# Leveraging Machine Learning and Big Data for Smart Buildings: A Comprehensive Survey

Basheer Qolomany, *Graduate Student Member, IEEE,* Ala Al-Fuqaha, *Senior Member, IEEE*, Ajay Gupta, *Senior Member, IEEE*, Driss Benhaddou, *Member, IEEE*, Safaa Alwajidi, Junaid Qadir, *Senior Member, IEEE*, Alvis C. Fong, *Senior Member, IEEE*

*Abstract*—Future buildings will offer new convenience, comfort, and efficiency possibilities to their residents. Changes will occur to the way people live as technology involves into people's lives and information processing is fully integrated into their daily living activities and objects. The future expectation of smart buildings includes making the residents' experience as easy and comfortable as possible. The massive streaming data generated and captured by smart building appliances and devices contains valuable information that needs to be mined to facilitate timely actions and better decision making. Machine learning and big data analytics will undoubtedly play a critical role to enable the delivery of such smart services. In this paper, we survey the area of smart building with a special focus on the role of techniques from machine learning and big data analytics. This survey also reviews the current trends and challenges faced in the development of smart building services.

## I. INTRODUCTION

Although the term "smart building" (SB) may bring a thought of a fictional smart space from science-fiction movies, but the reality is that SBs exist today, and their number is getting increased. With recent advances in machine learning (ML), big data analytics, sensor technologies and the Internet of Things (IoT), regular buildings can be cost-effectively transformed into SBs with bare minimum infrastructural modifications. There are smart office, smart library, smart home, smart health care facilities, smart hospital and many other types of SBs that can provide automated services that can provide many value-added services (such as reduction of wasted energy) and also help to ensure the comfort, health, and safety of the occupants.

The hyperconnectivity that will be brought about by the emergence of IoT will increase the promise of SB since now all the basic building amenities and commodities ranging from your house electronics to your plant vases will be interconnected. But this hyperconnectivity will at the same time complicate the process of managing SBs. In particular, SBs and their inhabitants are expected to create large volumes of streaming data. ML, sampling, compression, learning, and filtering technologies are becoming more significant to manage the stream of big data of individuals.

many other types of SBs

In 1981, the term Intelligent Buildings (IBs) was initially coined by United Technology Building Systems (UTBS) Corporation in the U.S. In July 1983, IBs became a reality with the opening of the City Place Building in Hartford, Connecticut [1]. Today, the number of SBs is growing at an unprecedented rate including smart office, smart hospitality, smart educational facilities etc. [2]. An SB is recognized as an integrated system that takes advantage of a range of computational and communications infrastructure and techniques [3]. Examples of SB services include smart thermostats that allow the temperature to be controlled based on the time of the day/year and the users' preferences with minimal or no manual configuration. Using data analytics to "learn" the users' preferences before taking the appropriate actions is probably the most important enabling technology for IBs [4]. Lately, smart coffee machines appeared in the market with the capability to make coffee automatically, according to users' preferences and schedules. Fridges can offer allocated programming interfaces for their control [5]. IBs aim to provide their users with safe, energy efficient, environment-friendly, and convenient services.

In order to maximize comfort, minimize cost, and adapt to the needs of their inhabitants, SBs must rely on sophisticated tools to learn, predict, and make intelligent decisions. SB algorithms cover a range of technologies, including prediction, decision-making, robotics, smart materials, wireless sensor networks, multimedia, mobile computing, and cloud computing. With these technologies, buildings can cognitively manage many SB services such as security, privacy, energy efficiency, lighting, maintenance, elderly care, and multimedia entertainment.

The massive volume of sensory data collected from sensors and appliances must be analyzed by algorithms, transformed into information, and minted to extract knowledge so that machines can have a better understanding of humans than their environment. Furthermore, and most importantly, such knowledge can lead to new products and services that can dramatically transform our lives. For example, readings from smart meters can be used to better predict and balancing the usage of power. Monitoring and processing sensory data from wearable sensors attached to patients can produce new remote

B. Qolomany, A. Al-Fuqaha, A. Gupta, S. Alwajidi, and A. Fong are with the Department of Computer Science, Western Michigan University, Kalamazoo, MI 49008 USA (e-mail: basheer.qolomany@wmich.edu; ala.al-fuqaha@wmich.edu; ajay.gupta@wmich.edu; safaakhalilmu.alwajidi@wmich.edu; alvis.fong@wmich.edu).

D. Benhaddou is with Engineering Technology Department, University of Houston, Houston, Texas 77204 USA (e-mail: dbenhadd@central.uh.edu).

J. Qadir is with Information Technology University, Lahore, Pakistan (e-mail: junaid.qadir@itu.edu.pk).

healthcare services.

The main philosophy behind ML is to create the analytical models automatically in order to permit the algorithms to learn continuously from available data. The application of ML techniques increased over the last two decades due to the availability of massive amounts of complex data and the increased usability of current ML tools. Today, ML is already widely applied in different applications including recommendation systems offered by online services (e.g., Amazon, Netflix) and automatic credit rating services used by banks. Alphabet's Nest thermostat utilizes ML to "learn" the temperature preferences of its users and adapt to their work schedule to minimize the energy use. Other widely publicized examples of ML applications include Google's self-driving car, sentiment analysis of Amazon and Twitter data, fraud detection, and Facebook's facial-recognition technology that is used to tag the suggested person on images uploaded by users.

*A. SB Trends and Market Impact*

In this section, we look at the statistics related to SBs, to allow us to understand the current trends and motivations in industry marketplaces and academic researches toward SBs.

According to the report by MarketsandMarkets [6], The SB market is estimated to grow from 7.42 billion dollars in 2017 to 31.74 billion dollars by 2022, at a Compound Annual Growth Rate (CAGR) of 33.7% from 2017 to 2022. In yet another report Zion Market Research [7], 2016 and it is expected to reach 61,900 million dollars by 2024. It is expected to exhibit a CAGR of more than 34% between 2017 and 2024. The market is primarily driven by government initiatives globally for SB projects and the increasing market for integrated security and safety systems as well as energy efficient building systems. Figure 1 shows the Statista [8] forecast market size of the global smart home market from 2016 to 2022 (in billion U.S. dollars).

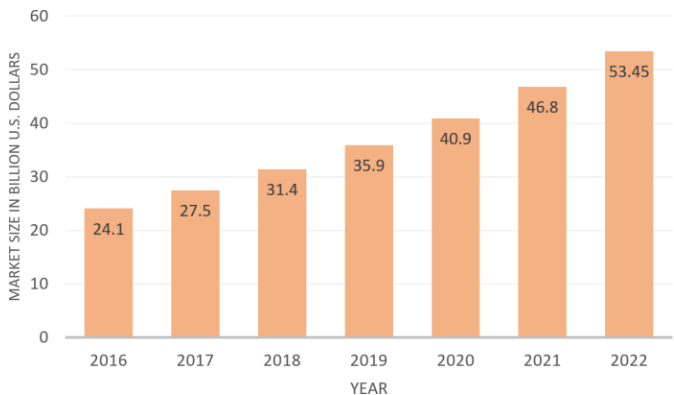

Fig. 1: Forecast market size of the global smart home market from 2016 to 2022 [8].

According to the Gartner report [9], it is expected that the number of smart connected homes grows to 700 million homes by 2020, supplied by mass consumer adoption and an increase in the number of devices and apps available. Figure 2 shows

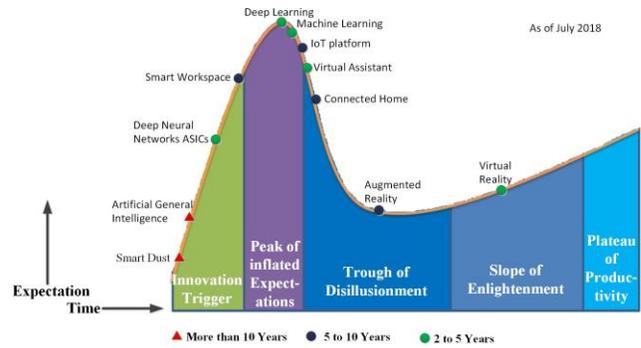

Fig. 2: Hype Cycle for the Connected Home, 2018 [7].

Gartner's 2018 Hype Cycle expectation for deep learning, ML, connected homes, and smart workspace.

According to report by Research and Markets [10][11], the global IoT SB market will reach approximately $51.44B USD globally by 2023. The report also forecast that 33% of IoT SB market will be supplied by artificial intelligent technologies by 2023, and automation systems of SB will grow at 48.3% CAGR from 2018–2023. Frost & Sullivan also predict that by 2025, the growth of connected home living will reach 3.7 billion smartphones, 700 million tablets, 520 million wearable health-related devices and 410 million smart appliances in the connected person world.

*B. Related Survey Papers*

Although many of survey papers focused on SBs have been published, none of them is focused on the role of data analytics and ML in the context of SBs. We describe the relevant survey papers next and will compare these survey papers to our paper in Table II.

- Chan et al. in 2008 provided an overview of smart home research [12]. It also discusses assistive robots, and wearable devices. The article reviews smart home projects arranged by country and continent.
- Alam et al. [13] provided details about sensors, devices, algorithms, and communication protocols utilized in smart homes. The paper reviews smart home works according to their desired services and research goals; namely, security, comfort, and healthcare.
- Lobaccaro et al. [14] presented the concept of smart home and smart grid technologies and discuss some challenges, benefits and future trends of smart home technologies.
- Pan et al. [15] reviewed the works on efficient energy consumption in SBs using microgrids. The survey investigates research topics and the recent advancements in SBs and the vision of microgrids.
- A few survey papers have reviewed works on facilitating independent living of the elderly people in smart homes. Ni et al. [16] conducted a survey on the features of sensing infrastructure and activities that can assist the independent living of the elderly in smart homes. A survey on ambient assisted living technologies for elderly people has been presented Rashidi and Mihailidis [17]. Peetoom et al. [18] focused on monitoring technologies to recognize life activities in-home such as fall detection and changes in health status. Salih et al. [19] presented a review of ambient intelligence assisted healthcare monitoring services and described the various application, communication, and wireless sensor network technologies that have been employed in the existing research literature.

- A number of papers have focused IoT: (a) Perera et al. [20] discussed IoT applications from the perspective of context-awareness and self-learning; (b) Tsai et al. [21] surveyed the applications of data mining technologies in IoT; and (c) Mahdavinejad et al. [22] reviewed some ML methods that can be applied to IoT data analytics.

*C. Contributions and Organization of This Paper*

To the best of our knowledge, this is the first survey that covers SBs jointly from the perspectives of application, data analytics, and ML. The main contributions of our paper are:

- Exploration of the potential of ML-based context-aware systems to provide SB services;
- Identification of research challenges and directions for SBs and how ML models can help in resolving such challenges;
- Identification of SB applications including comfort, security, energy efficiency, and convenience and the role of ML in such applications. Our research can provide an impetus to ML researchers to investigate new exciting ML-based SB services.

The rest of the paper is organized as follows: Section II introduces the concept of SBs and its underlying architecture. Section III introduces the various components of the SB ecosystem and its underlying architecture. Section V presents context recognition and activity modeling and the role of ML in SBs. Section VI highlights research and development challenges and provides a future perspective of SB projects. Finally, Section VII presents a summary of lessons learned and concludes the paper.

For the convenience of the readers, we have enlisted the important acronyms used in Table I.

II. SMART BUILDINGS: CONCEPT AND ARCHITECTURE

In 1984, The New York Times published an article that described that real estate developers are creating "*a new generation of buildings that almost think for themselves called intelligent buildings*." Such an intelligent building (IB) was defined as "a marriage of two technologies old-fashioned building management and telecommunications." [23]. Since then, many definitions of SBs have been suggested. This is due to the fact that the life-cycle of building planning, design, implementation, and operation involves different industry players that have different roles. In addition, the rapid changes in technology are affecting this definition. For instance, the advent of IoT and smart city concepts is impacting the definition of SB. Therefore, it is hard to compose a unique view of IBs with a single definition that is accepted worldwide. However, it is vital to have a good understanding of the main standard bodies and companies involved in shaping the development of SBs [1]. The Institute for Building Efficiency [24] focuses on the operation of buildings to provide efficient healthy and comfortable environment [25]. IBM [26] focuses also on the operation of SBs to provide integrated physical and digital infrastructures that provide reliable, sustainable, and cost-effective occupancy services. According to the European Commission's Information Society [27], SBs means buildings that are supplied by information and communication technologies in the context of the combining Ubiquitous Computing and the IoT: In general, the buildings that are supplied with sensors, actuators, microchips, micro- and nano-embedded systems in order to enable collecting, filtering and producing more information locally, to be further incorporated and managed globally according to business functions." In SBs, a variety of AI and multi-agent system techniques are employed including [28]:

1) Reasoning and knowledge representation including ontologies and rules to represent devices and home services.
2) ML for human activity recognition.
3) Multi-agent systems for distributed intelligence and semantic interoperability.
4) Intelligent approaches such as planning, intelligent control, adaptive interfaces, and optimization for efficient management of resources and services.

TABLE I: LIST OF IMPORTANT ACRONYMS USED

| Acronym | Definition |
|---|---|
| AAL | Ambient Assisted Living |
| ANNs | Artificial Neural Networks |
| AODE | One-Dependence Estimators |
| APAC | Asia and Pacific |
| AR | Accuracy Rate |
| BBN | Bayesian Belief Network |
| BT-LE | Bluetooth Low Energy |
| CAGR | Compound Annual Growth Rate |
| CAN | Controller Area Network |
| CART | Classification and Regression Tree |
| CEA | Consumer Electronics Association |
| CEP | Complex Event Processing |
| CHAID | Automatic Interaction Detection |
| CNN | Convolutional Neural Network |
| CNN | Convolutional Neural Networks |
| CRF | Conditional Random Field |
| DBM | Deep Boltzmann Machine |
| DBN | Deep Belief Networks |
| DIY | Do-It-Yourself |
| ECG | Electrocardiography |
| EEG | Electroencephalography |
| EM | Expectation Maximization |
| EMG | Electromyography |
| EMSs | Energy Management Systems |
| EOG | Electrooculography |
| ET-KNN | Evidence Theoretic Knearest Neighbors |
| FLS | Fire and Life Safety |
| GBM | Gradient Boosting Machines |
| GBRT | Gradient Boosted Regression Trees |
| GSR | Galvanic Skin Response |
| GUI | Graphical User Interface |
| HDFS | Hadoop Distributed File System |
| HMMs | Hidden Markov Models |
| HVAC | Heating, Ventilation, and Air Conditioning |
| IBs | Intelligent Buildings |
| IBT | Intelligent Building Technology |
| ICA | Independent Component Analysis |
| ICT | information and communication technologies |
| ID3 | Iterative Dichotomiser 3 |
| IoT | Internet of Things |
| ISM Bands | Industrial Scientific Medical Bands |
| KNX | Konnex |
| LANs | Local Area Networks |
| LCR | Lighting Control and Reduction |
| LDA | Linear Discriminant Analysis |
| LOESS | Locally Estimated Scatterplot Smoothing |
| LoT | Lab of Things |
| MARS | Multivariate Adaptive Regression Splines |
| M-Bus | Meter-Bus |
| MISs | Management Information Systems |
| ML | Machine Learning |
| NFC | Near Field Communication |
| NLP | Natural Language Processing |
| OA | Office Automation |
| OLSR | Ordinary Least Squares Regression |
| ORE | Oracle R Enterprise |
| OSX | Oracle Stream Explorer |
| PCA | Principal Component Analysis |
| PCR | Principal Component Regression |
| PLC | Powerline Communication |
| PSNR | Peak-Signal-to-Noise Ratio |
| RBFN | Radial Basis Function Network |
| RBM | Restricted Boltzmann Machine |
| RDDs | Resilient Distributed Datasets |
| RFID | Radio Frequency Identification |
| RNN | Recurrent Neural Networks |
| SB | Smart Building |
| SVMs | Support Vector Machines |
| TMSs | Temperature Monitoring Systems |
| UTBS | United Technology Building Systems |

TABLE II: COMPARISON OF RELEVANT SURVEY PAPERS

| Cite | Purpose | Limitations |
|---|---|---|
| Chan et al. [12] | Review SH projects arranged by country and continent as well as the associated technologies for monitoring systems and assistive robotics | Does not focus on the role of ML and big data analytics, it does not review and categorize the papers according to the applications of SH |
| Alam et al. [13] | Reviews SH projects according to research objectives and services; namely, comfort, healthcare, and security. | Does not focus on the role of ML and big data analytics for SB. |
| Lobaccaro et al. [14] | review of existing software, hardware, and communications control systems for SH and smart grid | Does not focus on the role of ML and big data analytics. It also does not focus on reviewing and categorizing papers according to the applications of SH. |
| Pan et al. [15] | Review the research topics on the energy efficiency and the vision of microgrids in SBs. | The focus of the paper is not the ML and big data analytics for SB services. It also does not cover other applications of SB rather than energy efficiency. |
| Ni et al. [16] | propose a classification of activities considered in SH for older peoples independent living, they also classify sensors and data processing techniques in SH | Does not cover all the services in SH. It also does not categorize the research according to different ML model styles. |
| Rashidi and Mihailidis [17] | Review AAL technologies, tools, and techniques | The paper focuses only on AAL in healthcare, and does not cover the other applications in SH or SB; in addition, there is no classifying of the researches according to ML model styles |
| Peetoom et al. [18] | Review the works on monitoring technologies that detect ADL or significant events in SH. | Does not focus on the role of ML in SB. |
| Salih et al. [19] | Review the works on ambient intelligence assisted healthcare monitoring focuses only on AAL in healthcare, and does not cover the other applications in SH or SB. | The paper also does not show the challenges and the future research directions in the field. |
| Perera et al. [20] | Review the works in context awareness from an IoT perspective | Does not focus specifically to the SB domain and its application services. |
| Tsai et al. [21] | Review the research works of data mining technologies for IoT applications. | Does not focus specifically to SB applications. |
| Mahdavinejad et al. [22] | Review some ML methods applied to IoT data by studying smart cities as a use case scenario. | Does not focus on SB and its applications as a use case. |

An SB is therefore the integration of a wide range of systems and services into a unified environment that involve energy management systems, temperature monitoring systems, access security systems, fire and life safety, lighting control and reduction, telecommunications services, office automation, computer systems, area locating systems, LANs, management information systems, cabling and records, maintenance systems, and expert systems [29].

Figure 3 shows examples of SB appliances including air-conditioning systems, lighting systems, solar energy generators, power-supply systems, temperature sensors, humidity sensors, power usage sensors, and surveillance cameras. For example, centralized control of these elements can promote the efficient use of energy through the intelligent control of lights and air conditioning units and the intelligent management of multiple green and brown energy sources. In most cases, an SB uses an Ethernet backbone with bridges to a Controller Area Network (CAN) [26].

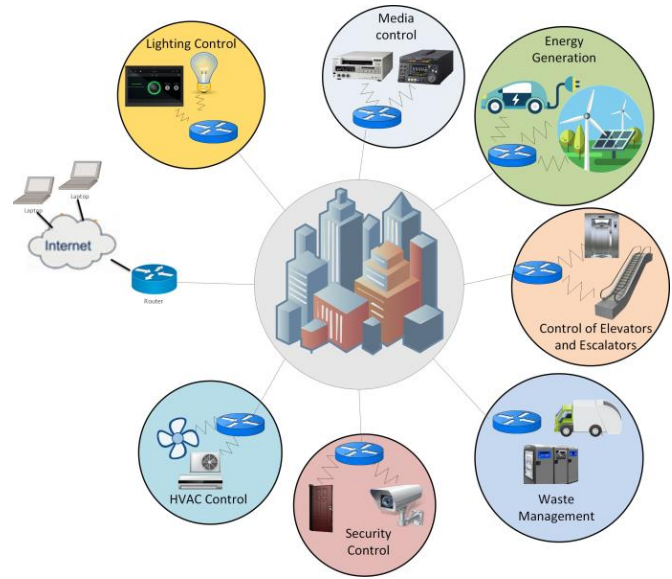

Fig. 3: Example of SB appliances.

It is easier to introduce smart services in residential buildings compared to commercial buildings since residential buildings have less technical equipment and less stringent efficiency requirements. Because the commercial buildings usually have more public visitors and therefore building models for com-mercial buildings are usually more challenging than building models for residential buildings which usually have a limited number of the occupants most of the time. In addition, the costs associated with the purchase and installation of smart devices and infrastructure at commercial buildings is more than residential buildings. Figure 4 shows an integrated framework in a residential building that employs a network of intelligent sensors. These sensors control systems such as energy generation, metering, HVAC, lighting, and security. A building automation system manages a set of smart appliances, sensors, and actuators, which collectively deliver services for the well-being of the inhabitants. Examples of such smart appliances, sensors, and actuators include washers and dryers, refrigerators, heaters, thermostats, lighting systems, power outlets, energy meters, smoke detectors, televisions, game consoles, windows/door controllers and sensors, air conditioners, video cameras, and sound detectors. More advanced smart devices are constantly being developed like smart floors and smart furniture [28], [30].

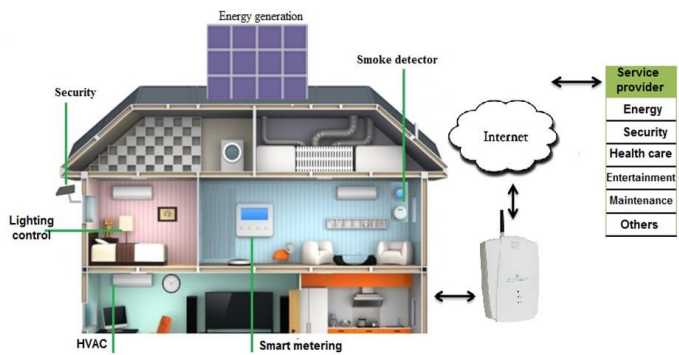

Fig. 4: Smart appliances, sensors, and actuators in a smart residential building.

The IoT will enable the integration and interoperability of heterogeneous devices in SBs as well as the real-time processing of the data generated by sensors in support of optimal control and operation of the building. IoTs are based on an architecture modeled in layers as depicted in Figure 5. As can be seen from the *sensing layer* (the bottom layer in Figure 5), input data is obtained from different types of physical sensors that monitor environmental parameters, collect data about residents and detect anomalies (e.g., fire and water pipe bursts). This layer also includes actuators that can be controlled to save energy, minimize water consumption, etc.

The *network layer* (the second layer in Figure 5), includes access and core networks that provide transparent data transmission capability. This layer serves as a bridge between the sensing layer and the upper layers which are mainly responsible for data processing.

An intermediary software layer called the *middleware layer* is needed (the third layer in Figure 5) to provide seamless integration of heterogeneous devices and networks covered by the sensing layer of the architecture. That layer serves as a bridge between the embedded software that runs of smart sensors and back-end software services. This layer provides interoperability using standardized programming interfaces and protocols [31]. Therefore, this layer performs the process of converting the collected data from various data formats into a common representation. SB middleware can be based on open standards or proprietary, in addition, application-specific or general-purpose. Most often, proprietary middleware is application-specific while general-purpose middleware is based on open standards [28].

The *context and semantic discovery layer* (the fourth layer in Figure 5) is responsible for managing context and semantic discoverers including context and semantics generating, configuring, and storing.

The *processing and reasoning layer* (the fifth layer in Figure 5) is responsible for processing the extracted information from the middleware then according to the applications type it will make decisions. In this layer, there are various techniques of information processing applied to fuse, extract, contextualize. massive data into useful actionable knowledge. In this layer, two phases should be identified: context consumer and context producer of the middleware. In the context consumer phase, the data processing techniques are applied on the data produced by the middleware; while in context producer phase the process of decision-making is implemented to supply the service layer with valuable knowledge. while in the second stage, further context information can be provided to the middleware for registration in the ontology context.

Specific services and applications are abstracted in the *application layer* (the top-most layer in Figure 5). This layer presents a framework with direct access to the underlying functionalities to serve in the implementation of various types of applications. Moreover, control panels should be installed in the building to control the automated indoor spaces and to support a local human-machine interface. For instance, in a multi-story building, each floor could have a control panel to automate the operations, such as control opening the windows, control of air conditioning to achieve the desired temperature, control close/open the blinds according to the preferred light intensity before using artificial lighting. [32], [33].

Summary: Still there is no single standard definition for SBs. In this section, we reviewed many definitions for SBs by many institutes, counties, regions and different disciplines; each has their own definition for SBs. We presented the layered architectural pattern for adapting services in an SB environment. We wanted to provide a general design for adapting actions according to the different versions of context in SBs. This architecture may be used in different smart environments such as intelligent transport systems, security, health assistance, and SBs among others. We layered the architecture into six layers starting from the sensing layer, which includes various types of sensors that are installed to collect environmental information in SBs. While network layer providing data stream support and data flow control and ensuring that messages arrive reliably by using data transport protocols such as Wi-Fi, Bluetooth, Ethernet etc. Data Acquisition layer to collect the data from the heterogeneous sources of data. Context and semantic discovery layer to generate, configure, and store context and semantic information. Context processing and reasoning layer to process

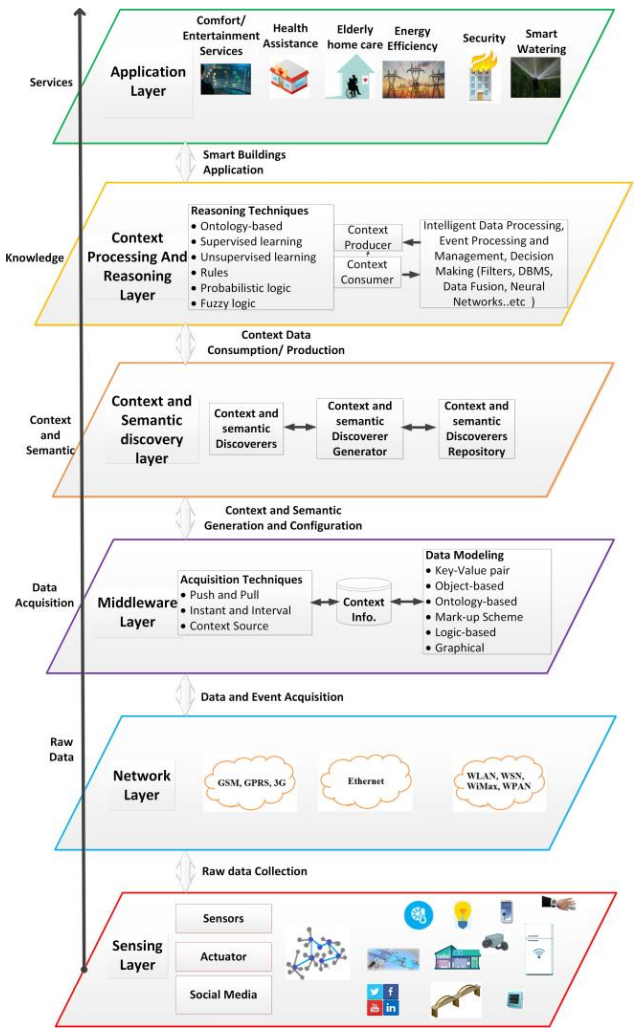

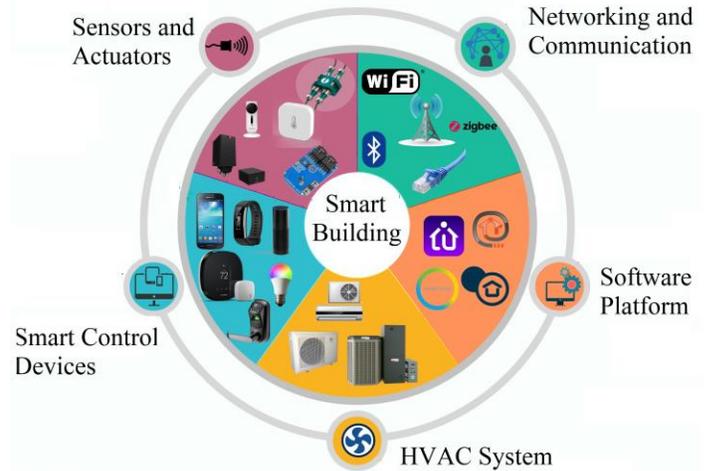

Fig. 6: Components of smart buildings.

Current systems utilize control devices and smart sensors that are connected to a central system. These control devices and smart sensors are placed throughout the environment. Each particular system has its own collection of networking and communication systems that enable it to communicate with the central system. SBs are performing connected networks that serve as a communication backbone for multiple systems. In many ways, HVAC equipment is the most complicated building system, with numerous components arranged to produce heating, cooling, and ventilation. The functionality of HVAC system not only makes the building healthy and comfortable for its inhabitants, but it also manages a big part of the energy consumed, as well as plays a significant role in life safety. SBs adopt technology to monitor and control facility systems and perform any required modifications. The objective of an SB is to utilize computers and software to control lighting, alarm systems, HVAC, and other systems through a single computer interface.

Fig. 5: Layers of the base IoT architecture that serves as the foundation for SBs.

the information and extract the knowledge that making the decisions according to the application context. And the last layer which is application layer such as health assistance and elderly home care, comfort and entertainment services, security, tele-management, smart watering, energy efficiency, etc. After discussing the main components of commercial and residential buildings, we have now set the stage for a detailed discussion on the components of SBs in the next section.

## III. SMART BUILDING COMPONENTS

Advances in smart building technology have driven to the extensive development of SBs to generate economic and environmental benefits for building owners through the convergence of IT and building automation systems. Figure 6 shows the key components of SB systems, these include extensive sensors and actuators systems, networking and communication systems, software platform system, HVAC system, and smart control devices.

### A. Sensors and Actuators for SBs

Sensors and actuators are mechanical components that measure and control the environmental values of their environment. Sensors collect information from the environment and make it ready for the system. For instance, IR sensors can be utilized for human presence detection in a room. While actuator is a device to convert an electrical control signal to a physical action, such that it takes decisions and then performs proper actions according to the environment, which enables automated and remote interaction with the environment. For example, a light actuator is capable of switching on/off, dimming one or more electric lights [34]. The rapid development of micromechanics, microelectronics, integrated optics, and other related technologies has facilitated the development of different types of smart sensors integrated into daily objects and infrastructure at smart building environment or worn by the users, and are connected by network technologies in order to collect contextual information about daily living activities more efficiently and faster, with lower energy consumption and less processing resources. Environmental sensors are utilized for detecting the human activity of a specific object that performed in specific locations in the building, while wearable sensors are utilized for controlling and observing mobile activities and physiological signals [35].

*1) Environmental Sensors:* It is found that data collected from environmental sensors can form important information to monitor human behaviors within an SB. These sensory data are then analyzed to identify and observe basic and instrumental daily living activities made by occupants such as bathing, dressing, preparing a meal, taking medication etc. The environmental sensing is generally based on several simple binary sensors in every part of the home, RFID technology, and video cameras. This variety of sensing may implement important insight into contexts and actual activities although it might come with possible costs such as complexity. Motion sensors are utilized for detecting the occupants presence and location everywhere in the house. There are different types of motion sensors. IR presence sensor is one of the most utilized kind of motion sensors in SBs to detect occupants presence. Pressure sensors can be attached to the objects such as beds, chairs, sofas, and floors in order to track the actions and locations of the occupants. While Contact switches are usually placed on the doors of fridge, rooms, or cabinets to detect the actions that the occupant makes with these objects [36]. Light sensors, humidity sensors, temperature sensors, or power sensors are other types of sensors that are deployed and utilized in SB to recognize the activities. Light sensors are utilized to measure the light intensity in a particular room in the building. Humidity sensors are utilized to detect the air humidity of a specific location in the building. Temperature sensors are utilized to measure the temperature of the specific environment. while the power sensors are utilized to identify the power usage of electric devices.

*2) Wearable Sensors and Biosensors:* These sensors are attached directly or indirectly to the user body. Their small size enables these sensors to be attached to clothes, wristwatches, glasses, belts, shoes etc. These sensors can be categorized into inertial sensors and vital sign sensors (or biosensors). Wearable inertial sensors are highly transportable and no stationary units that can give accurately detailed features of occupant's action and body posture. Those sensors are composed of accelerometers, gyroscopes and magnetic sensors. There is a need for receivers and cameras in the process of data collection, therefore can be used outside laboratory circumstances [37].

wearable biosensors such as blood pressure, skin temperature, and heart rate are significant for collecting vital signs to monitor the health. The most commonly utilized inertial sensors for mobile activity monitoring are accelerometers and gyroscopes. Accelerometers can be utilized to measure the rate of acceleration accompanying a sensitive axis, they are useful to monitor the motion's activities such as doing exercise, standing, sitting, walking, or walking upstairs and downstairs. While the gyroscopes can be utilized to measure angular velocity and maintain orientation. Some examples of primary vital signs are Electrocardiogram (ECG), heart rate, blood pressure, blood glucose, oxygen saturation, and respiratory rate. There are various vital sign sensor utilized to measure different vital signals such as Electroencephalography sensors (EEG) for observing electrical brain activity, Electrooculography sensors (EOG) for observing eye movement in ocular activity, Electromyography sensors (EMG) for observing muscle activity. Electrocardiography sensors (ECG) for observing cardiac activity, pressure sensors for observing blood pressure, $CO_2$ gas sensors for observing respiration, thermal sensors for observing body temperature and galvanic skin response for observing skin sweating [38][39].

*3) Heating, Ventilation, and Air Conditioning (HVAC):* HVAC system plays an essential role in SB services. HVAC system plays a remarkable role in efficient energy consumption in SBs, as well as it offers new operating options to increase the occupants' comfort. In addition to meeting the desired temperature, HVAC control systems are produced in order to sustain comfort within an enclosed space by producing a specific level of humidity, pressure, air motion, and air quality in an SB [40]. $CO_2$, humidity and temperature levels in a building can affect occupant's health and comfort; consequently measuring $CO_2$, humidity, and temperature in this context can improve personal wellbeing [41]. Heating and cooling systems consume a huge amount of energy in the buildings, so it is necessary to optimize it utilizing smart controllers and sensors in order to save operational costs. Smart HVAC systems can sense and control efficiently different air quality parameters inside the building by utilizing distributed sensors and VAV fans throughout the building to perform an optimal ventilation [42]. Most of the current HVAC actuation systems in smart buildings are based on the data collected about the occupants using sensors and cameras, which are utilized specifically for HVAC systems. Certainly, There is a specific cost for the design, maintenance, setup and hardware of the data collection network [43]. Table III shows a summary for different types of smart sensors in the SBs.

## B. Smart Control Devices

Smart control devices collect data from a variety of sensors, process this data, and activate actuators to react to the events detected by the sensors. A smart control device can operate independently, without control by a central server. But there might be a needed communication amongst various control devices or they can connect with each other using the smart gateway.

*WeMo* [44] is a Wi-Fi enabled switch utilized to turn electronic devices on/off from anywhere. It can control LED motion sensors, light bulbs, mart wall switches and plugs, and lighting devices, all from the smartphone app or browser. There is no hub needed for WeMo devices, everything can be managed through the free cloud service provided by Belkin. You can use the specific channel to connect the device to e-services such as Gmail to trigger specific actions. WeMo devices also support context-aware feature, it turns on/off automatically according to the time of day, whether it is sunrise or sunset etc.

The *Nest* thermostat [45], a smart device developed by Nest—which has been acquired by Google—adjusts to your life and seasons change automatically. Just use it for a week and it programs itself. It learns about the level of temperatures that the occupants prefer and creates a context-aware personalized schedule. The smart thermostat turns to an energy-efficient mode automatically when the residents leave the building. It could start warming up the area when it senses activity, such as an occupant's returning back home from work. The Nest Thermostat is controllable via a smartphone and an installed app. If you are away for a while, this device has also a capability to sustain a particular temperature in your house.

TABLE III: VARIOUS SMART SENSORS USEFUL IN THE CONTEXT OF SBs

| Sensor | Measurement | Category |
| --- | --- | --- |
| Infrared sensor | User presence in a room | Environmental sensors |
| Video cameras | Human actions | Environmental sensors |
| RFID | Object identification | Environmental sensors |
| Motion sensor | Object/User presence/ location | Environmental sensors |
| Contact switch | Detect users' interaction with the object | Environmental sensors |
| Pressure sensor | Tracking movements and location of the user | Environmental sensors |
| Light sensor | Intensity of light | Environmental sensors |
| Temperature sensor | Temperature of surrounding environment | Environmental sensors |
| Humidity sensor | Detect the air humidity in a specific area | Environmental sensors |
| Power sensor | Detect the usage of electric devices | Environmental sensors |
| Accelerometer | The rate of acceleration accompanying a sensitive axis | Wearable inertial sensors |
| Gyroscope | Angular velocity and maintain orientation | Wearable inertial sensors |
| Electroencephalography | observing electrical brain activity | Wearable vital sign sensors |
| Electrooculography | observing eye movement of ocular activity | Wearable vital sign sensors |
| Electromyography | observing muscle activity | Wearable vital sign sensors |
| Electrocardiography | observing cardiac activity, pressure sensors for observing blood pressure | Wearable vital sign sensors |
| $CO_2$ gas sensors | observing respiration | Wearable vital sign sensors |
| Thermal sensors | observing body temperature | Wearable vital sign sensors |
| galvanic skin response | observing skin sweating | Wearable vital sign sensors |

*Lockitron* [46] is a door lock that can control the door remotely over the Internet to open and close it by phone. Lockitron app can be installed and used by any iOS or Android smartphone. Homeowners can directly grant family and friends the access to open a given door by providing authorization over the Internet. Lockitron can also utilize Bluetooth low energy technology, which means that it will keep controlling even in the event of Internet or power outages. Lockitron can also connect to the Internet with Bridge, through which occupants can control the bolt anywhere in the world.

The *SmartThings* [47] SB automation system comprised of a communications smart hub, that supports various smart appliances and devices; the smart hub supports various technologies and protocols such as ZigBee, Z-Wave, as well as IP-accessible devices and lets you control appliances using Wi-Fi and Bluetooth connectivity. SmartThings provides kits that include smart plugs, in addition, the basic sensors that can be utilized to measure temperature, as well as to detect presence, motion, orientation, and vibration. SmartThings also includes an open platform that enables smart device vendors and third-party software to provide hardware and software that can be utilized alongside the platform.

*Philips Hue* [48] is a combination of LED lighting with mobile technology. An accompanying mobile app that allows you to control lighting systems and changing color sets depending on your mood utilizing Wi-Fi technology. The new Philips Hue bridge supports the required authentication to enable Apple HomeKit technology to control and enable your Philips Hue to connect to other HomeKit enabled accessories and take control of your home.

*Blufitbottle* [49] this bottle records the drinking habits of the users and sends them notifications about the time and amount of the water that they are supposed to drink to keep them healthy and hydrated. The app collects data about users such as their weight and age, plus other factors such as the current levels of temperature and humidity to estimate the amount of the needed water to keep them hydrated. When the user falls behind with hydration, an alert sounds, as well as a simple glance from the LEDs, will indicate when it's time for the next drink.

*Canary* [50] is an all-in-one home security system that comprises a set of sensors such as temperature, air quality, sound, motion vibration, in addition, an HD video camera in one unit. The system utilizes ML algorithms to let the users know what is happening at home and take action by sending notifications to your phone if something happens. Those ML models learn over time and send the users smarter notifications as it detects motion. So that, the longer you have the system, the more effective it becomes. Canary is able to decrease the rate of false alarms by learning the user behavior and the ambient noise level and the home temperature patterns.

*Amazon Echo* [51] is a small cylinder enable the users to control anything in the home via the voice. Amazon Echo has a powerful voice recognition capability, the user does not have to worry about the complexities of their voice. Amazon Echo is connectable via Wi-Fi or Bluetooth, the users can send voice commands to control the speakers as well as other compatible devices such as Belkin's WeMo and Philips Hue. It can also use Amazon cloud Lambda service to send commands. To send any command It requires to include the name of the program, for instance, "Alexa, turn on TV". It also includes a network to distant servers, which slows down the response time.

*Honeywell Total Connect Remote Services* [52] this device merges personal smart home automation with security monitoring task. It enables the occupants to control and monitor everything in the home from lighting and window shades systems to security cameras and smoke alarms. the user can utilize a smartphone app or desktop-mounted hardware console for controlling and monitoring. It can provide real-time alerts, GPS vehicle and asset tracking, video viewing, and mobile control. The system only supports Z-Wave devices, it needs to be installed by an authorized Honeywell dealer. It does not work with Wi-Fi enabled smart thermostats. In addition, the Honeywell provides security cameras and sensors, it also supports other smart devices from third parties, such as Yale locks and Lutron lighting. Table IV shows a comparison among various smart control devices in the SBs.

TABLE IV: COMPARISON AMONG VARIOUS SMART CONTROL DEVICES IN SB

| | Technology | Platform used | Pros | Cons |
|---|---|---|---|---|
| WeMo | Wi-Fi | Android, iOS, Window phone | Affordable hardware options. Can expand using SmartThings hub. Can expand using IFTTT. | No color bulbs, no dimmer switches. Experienced some latency issues. |
| Nest Thermostat | Wi-Fi, Zigbee, Thread | iOS, or Android, Mac OS, Windows | Easy to program. It learns the user daily routine, it could set itself up for the user lifestyle after the first week of use. | More expensive than other smart thermostats. The Nest might not be for you if you are a stickler for temperatures that are "just right". |
| Lockitron | Bluetooth | Android, iOS | Affordable, easy to install, quiet operation, Offers proximity locking and unlocking. | Wi-Fi bridge costs extra. Does not work with other devices. |
| SmartThings | Wi-Fi, Zigbee, Z-Wave, Bluetooth | Android, iOS, Window phone | Affordable. Easy to install. Quiet operation. Offers proximity locking and unlocking. | Compatibility issues with other devices, workarounds for non-natively supported devices will be difficult for some. |
| Philips Hue | Wi-Fi | Android, iOS | Modest list of 3rd party integrations. Offers own proprietary hardware. Active web community for help. Deep customization for power users. | Least user-friendly app. Complex to configure simple tasks. Missing some "key" integrations. |
| Blufitbottle | Bluetooth | iOS and Android device | Easy to set up and use. Excellent light quality | The Philips HUE kit, has a fairly high cost. |
| Canary | Wi-Fi | iOS or Android device | Simple and easy to set up. | Expensive. |
| Amazon Echo | Wi-Fi, Bluetooth | Fire OS, Android, and iOS | High-quality voice recognition. Integrations with all of the "key players". Works for all users, regardless of phone brand. | The system only detects the intruder once they are inside premises. |
| Honeywell | Wi-fi, Z-Wave | iOS and Android | Easy installation. | Priced higher than other smart thermostats with similar functionality. |

*C. Networking and Home Gateway*

An SB combines a communication network in order to control smart devices and services within the building. The communication network of a smart building can be based on diverse communication media such as twisted pair cable, as the traditional computer networks. The networking in building automation system has a tendency to utilize a heterogeneous network that is made up of diverse communication media and network standards. The building automation network is identified by physical technology and communication protocols. There is an internal network that connects devices inside the building, as well as the external networks, can be integrated separately. Public Internet, ISDN, and mobile phone networks are some examples of external networks. [53], [28].

A typical SB may comprise a number of different components, such as sensors, actuators, communication and processing devices. Because of their nature, these components have limited capabilities and computational capacity in term of battery capacity and capability of data processing. To deal with this issue, most of the SB systems have been utilized as a central gateway to collect, process, and analyze context data from different sensors and actuators in the building. Several protocols such as Bluetooth, ZigBee, Wi-Fi, and Z-wave can be utilized for communicating the gateway. The home gateway can also collect and store data for a specific time period. Typically, these gateways can connect to the cloud services and perform data processing and reasoning tasks. The centralized gateway usually does not have any interface. They can be controlled and managed utilizing smartphones, tablets, or computers [54], [55].

In general, depending on the communication media used, SB network technology can be classified by interconnection method into three main types: Powerline, Busline, and Wireless [15], [56], which we describe next.

*1) Powerline communication (PLC):* PLC method reuses the building electrical network; such that devices, appliances, and services are directly connected to the main power supply utilizing the already available electrical outlets in a building. The data is sent through the normal cable system to activate or deactivate the devices in the building. PLC system is historically the oldest technology in SB and is generally cheap but less reliable and scalable [3]. Originally, the application of PLC was mainly to secure the typical operation of the electric power supply system in case of failures or breakdowns through the direct exchange of information between the distribution center, and power plant. Therefore, this approach has become a competitive choice for SB networking, benefiting from availability, robustness, and ready connectivity of this method. Some of the protocols of this method offer a single-way communication, which enables the device to only receive information but not to communicate. There are different mainstream protocols of PLC method such as X-10, INSTEON, HomePlug, BACnet, and Lonworks.

*2) Busline:* Busline systems in SBs networks use a separate physical media, usually twisted-pair cabling similar to the physical cables utilized for network services for transporting electrical signals. This type of systems is pleasant the building's occupants, albeit the configuration process is complex, and it requires some knowledge of networking. Although the configuring complexity and installation cost of this system, the use of a separate cable could present a positive note about this approach, as it allows this method of networking to provide higher bandwidth, and to make it the most reliable of the three approaches. In addition, this technology usually supports a completed two-way communication protocol that enables the appliances to easily communicate with each other. [57]. Some of the protocols in Busline technology are Konnex (KNX), CAN (Controller Area Network), Modbus, Meter-Bus (M-Bus).

*3) Wireless interconnection:* Many of the new SB applications use wireless technologies such as infrared and radio frequency, which are more convenient for users due to their untethered nature and the elimination of cables. The devices within the smart building can communicate wirelessly as radio wave can penetrate through floors, cabinets, and walls [56]. Because of the complexity and cost of potential modifications and of the re-wiring process in a smart building, several different wireless technologies are rising to produce flexible networking patterns convenient to occupants without taking to consider the physical wiring and deployment of such wire in the building. Typically, there are various protocols for the wireless system such as Bluetooth, ZigBee, WLAN, Z-wave, RFID etc., which essentially work in the industrial scientific medical bands, particularly in the 2.4GHz frequency range. These wireless technologies are usually related to some control network concept in an SB such as low power consumption, high cost-effectiveness, low speed, flexibility in networking, deployment as well as building coverage [3]. The gateway is the central server of an SB that is commonly used in IoT solutions. The services provided by the gateway essentially concern to system management functionalities such as monitoring, controlling, and configuring the systems and their devices. It also supports some processing and data storage capabilities required for complex applications.

*D. Software platform*

For a building to be "smart," it is important that all the appliances and systems in the building communicate and exchange data securely with each other as well as with smartphones, tablets, and servers in the cloud. Software platforms play a critical role in exchanging, archiving and disseminating information through different protocols. These platforms use push, pull, publish/subscribe, etc. The goal of the joint commercial enterprises is to develop an open source software platform in order to make the process of data exchanging easier between the devices of different manufacturers. Therefore, the users will not have to worry in the future about the compatibility issues when utilizing electric and electronic devices of different manufacturers at home. In addition, the new platform can also offer a variety of different home services such as entertainment, energy efficiency, and security technology. Hence, this will enable creating different apps for these areas of use [58].

ABB, Robert Bosch GmbH, and Cisco Systems Inc. established an open-software platform called *Mozaiq Operations GmbH* [59] to unify smart home technology and offer interoperability across for all devices and services in the building, to simplify the experience for residents. It will enable users to seamlessly and intuitively customize their appliances and devices, regardless of manufacturers and brands of these devices, in order to improve energy efficiency and achieve a unique level of control and comfort. For instance, the user can close the blinds in the home either by a click from a smartphone or through a pre-set instruction; and switch off automatically all screen devices for the children to go outside to play. In a smart building, many devices and appliances can simply and securely share information with one another and with smartphones and other smart devices; and the Internet in general.

*Indigo Domotics* [60] is to implement the do-it-yourself smart building platform. Indigo home automation software controllers available for the Mac OS enables residents to combine an array of common INSTEON, Z-Wave and X10 devices for unparalleled control of your building lighting, sensors, thermostats, and appliances. With Indigo Touch (sold through iTunes app store, iOS only), users will easily achieve remote control of their appliances utilizing an iPhone, iPad, or iPod Touch. They also can use a web browser on any device to control their appliances virtually anywhere in the world. The users can receive texts or emails about specific events has been detected for doors opening/closing, power failures etc. Indigo, from Perceptive Automation, is the newest home automation software for the Mac.

*OpenHAB* [61] is an open-source software platform that follows a middleware approach for integrating different technologies in smart building systems into a single solution. OpenHAB platform address a variety of network technologies and appliances in the area of a smart building. Currently, the dependency on a particular vendor becomes a problem due to the lack of a common language that bridges the different devices with building automation system. The main goal of the OpenHAB platform is to integrate the new devices and technologies in a smart building system through a community-based approach. OpenHAB utilizes an OSGi based modular system for communicating between different technologies and devices. Bindings can be developed and deployed as an OSGi bundle to bridge a particular technology and device. There are different supported technologies exist such as EnOcean, KNX, Z-Wave, and others are supported through special bindings [62].

*SmartThings* [47] this platform composed of hardware devices, sensors, and software applications. Context information is collected from the sensors, this context is contributed to the reasoning and action that are performed by the application. For instance, the sprinkler in the garden can detect the rain, and switch itself off accordingly to save water. SmartThings kit comprises sensors, smart devices, and hub. While the SmartThings application is configured to enable users to control and monitor their building environment through a smartphone device. The SmartThings Hub works to connect the sensors, devices and building's internet router to one another and to the cloud. It is compatible with different communication protocols such as Zigbee, Z-Wave, and IP-accessible devices. In addition, the SmartThings is compatible with other sensors and devices such as thermostats, moisture sensors, motion sensors, presence sensors, locks and garage door openers [63].

*HomeOS* [64] is Microsoft's home operating system platform, that can be installed on a personal computer. It is an open platform that is not limited to Windows-based devices [65] [66]. with HomeOS platform, applications can be installed to maintain various context-aware functionalities, for example, taking an image by a door camera and sending it to the occupant when someone rings the doorbell. HomeOS provides a PC-like abstraction that manages and extends the technology of network devices to the users and developers in the smart building environment. Its design enables the users to map their protocol-independent services to support the applications with simple APIs, a kernel, and protocols of specific devices. HomeOS usually runs on an allocated computer such as a home gateway, it does not need any adjustments to commodity devices. HomeOS usually utilizes (i) Datalog-based access control to facilitate the process of managing technology in the smart home (ii) a kernel to incorporate the devices and applications and (iii) protocol-independent services to allow the developers manageable access to the devices.

*Lab of Things (LoT)* [67][68] is an experimental research platform that utilizes connected devices in the buildings. LoT offers a framework that provides deployment capabilities such as remote monitoring and updating of system health, and logging data collected from different appliances in cloud storage. It enables data sharing and collecting, sharing codes, connect hardware sensors to the software platform, and participants using HomeOS. The platform is designed to make it simple to design solutions that can be deployed in IoT based smart services such as healthcare, energy management services as it works in combination with HomeOS.

*Eclipse Smarthome* [69] is a framework that has a focus on heterogeneous environments such as smart building and ambient assisted living. This platform takes to consider a variety of existing communication mechanisms. Eclipse SmartHome works as an abstraction and translation framework that enables communications across system and protocol boundaries. It provides many relevant implemented extensions, protocols, and standards that are significant for smart building services. Those implementations can be of Java library or an OSGi bundle shapes so that they can be utilized independently from the rest of the project. The framework can work on different embedded devices such as a BeagleBone Black, an Intel Edison, or a Raspberry Pi. Extensions of Eclipse SmartHome are compatible with the solutions provided by different vendors. This means your code that is written for a specific purpose can be extended easily on commercial platforms. Eclipse SmartHome offers a variety of characteristics to allow you to design a special Smart Home solution for your expectations [70].

Apart from discussing various SB solutions, we will also highlight the popular simulator called Cooja is used widely by the research community to produce small simulations for relatively large wireless networks of embedding sensors and actuators; and connected devices, in order to develop, debug and evaluate systems based on the wireless sensor network technology. Cooja simulator is a Java-based wireless sensor network simulator. It is distributed with Contiki OS project. Cooja enables the emulation of the set of sensor nodes, in addition, it can simulate physical and application layers of the system [71]. There are three basic properties for the simulated node in Cooja: Its hardware peripherals, node type, and data memory. The node type can be shared among multiple nodes and defines properties that are common to all these nodes [72].

*Summary:* The field of SBs contains a variety of technologies, across commercial, industrial, institutional and domestic buildings, including building controls and energy management systems. Several organizations and institutions are working to supply buildings with technology that enables the residents to adopt a single device to control all electronic devices and appliances. In this section, we discussed the various components for SBs including sensors and actuators, smart control devices, smart gateway, networking and software platforms.

## IV. ML Background for SBs: Models, Tasks, and Tools

Massive data generated from sensors, wearable devices, and other IoT technologies provide rich information about the context of users and building status and can be used to design SB management. This context information is needed to extract useful and interesting insights for various stakeholders. When the data volume is very high, developing predictive models using traditional approaches does not provide accurate insight and we require newly developed tools from big data. Big data is primed to make a big impact in SBs and is already playing a big role in the architecture, engineering, and construction (AEC) industries [73], notably for waste analytics [74] and waste minimization [75].

ML is a powerful tool that facilitates the process of mining a massive amount of data that have been collected from different sources around us and make sense of a complicated world. ML algorithms apply a model on new data by learning the model from a set of observed data examples called a training set. For example, after being trained on a set of sample accelerometer data marked as walking or jogging, an ML algorithm can classify the future data points into walking and jogging classes. ML makes it relatively easy to develop advanced software systems without much involvement from the human side. They are applicable to many real-life problems in SB environments. One can also design and develop self-learning and collaborative systems.

ML does not remove the human element from data science—it draws on computers' strengths in handling big data to complement our understanding of semantics and context. It only needs training data to extract better features or parameters required to improve a given system. ML algorithms can be used to make predictions based on data patterns. It enables the computer to learn from the fed input data without being explicitly programmed so that ML algorithms can learn from and make predictions on input data [76][77]. Nest thermostat is an example of a device that applies a specific temperature in a specific room and at a certain time of day according to the occupants' preference. There are devices such as Amazon's Echo that can learn from voice patterns, and the others those learn from much more complex behavior and activity patterns.

## A. ML Models

ML techniques have been widely used to develop smart systems which can sense and react according to context modifications in SBs [78]. There are many different ML algorithms, according to the two well-known theorems No Free Lunch theorem and Ugly Duckling theorem. No Free Lunch theorem states "there are no algorithms that can be said to be better than any other", without prior information about the problem, any two algorithms may perform equally well in solving a problem. While Ugly Duckling theorem states "we cannot say that any two different patterns would be more similar to each other than any other pairs." [79].

Mainly, ML is categorized into four categories handling different types of learning tasks as follows: Supervised learning, unsupervised learning, semi-supervised learning and reinforcement learning (RL) algorithms Figure 7 shows ML styles. These categories are described next and a summarized comparison between these ML techniques is presented in Table V.

*1) Supervised Learning:* refers to developing algorithms based on a labeled training dataset, from which the learner should generalize a representation by building the system model that represents the relations between the input, output and system parameters. ML model is developed through a training process that continues on the input training data until the model reaches the desired level of accuracy [80], [81]. Some examples of common supervised ML algorithms are: naive Bayes model, decision tree, linear discriminant functions such as support vector machines (SVMs), artificial neural networks (ANNs), hidden Markov models (HMMs), instance-based learning (such as k-nearest-neighbor learning), ensembles (bagging, boosting, random forest), logistic regression, genetic algorithms, and logistic regression [82] [83]. Supervised learning approaches are extensively used to solve different problems in smart buildings.

*Application in SBs:* Boger et al. [76] proposed a supervised learning system using Markov decision processes to help people with dementia the process of hand washing. Altun et al. [84] make a comparative study on the supervised human activity classification approaches using body-worn miniature inertial and magnetic sensors. Mozer [85] developed the occupant comfort control of the home environment system using neural networks and reinforcement learning to control air heating, lighting, ventilation, and water heating in the smart home environment. Bourobou et al. [86] presented a hybrid approach using ANN and K-pattern clustering to identify and predict user activities in the smart environments. Hsu et al. [87] proposed a TV recommendation system using a neural network model based on user personalized properties such as activities, interests, moods, experiences, and demographic information data. Fleury et al. [88] proposed a healthcare-focused smart home system using the SVM algorithm to classify daily living activities based on the data from the different sensors.

Supervised learning problems can be further grouped into classification, regression, time series, and ensemble method problems.

*a) Classification:* The task of classification algorithms is to classify an instance into a specific discrete set of possible categories. Given two sets of data (labeled and unlabeled datasets), the labeled dataset is used for the training process, while the unlabeled dataset will be used to evaluate of the classification results. The normal process is to count the number of instances that are assigned to the right category, which is also known as the accuracy rate (AR) defined by [21].

The classification algorithm can mathematically be described as follows:

$$AR = \frac{N_c}{N_t} \quad (1)$$

where $N_c$ denotes the number of test instances that are correctly assigned to their categories to which they belong; $N_t$ the number of test instances. The precision (*P*) and recall (*R*) are used to measure the details of the classification results. The four possible outcomes are true positive (*TP*), false negative (*FN*), false positive (*FP*), and true negative (*TN*), the precision (*P*) and recall (*R*) are generally defined as:

$$P = \frac{TP}{TP + FP} \quad (2)$$

Given *P* and *R*, a simple method to describe the precision and recall of the overall classification results, called F-score or F-measure, is defined as:

$$F = \frac{2PR}{P + R} \quad (3)$$

Commonly used classification techniques include decision trees, SVM, rule-based induction, neural networks, deep learning, memory-based reasoning, and Bayesian networks [89].

*b) Decision Tree Algorithms:* The decision tree method is an important predictive ML modeling approach, which constructs a model of decisions presented based on the actual values of features in the data. Decision trees can be utilized for both classification and regression problems. In tree structures, leaves represent class labels and branches represent conjunctions of attributes that drive to those labels. [90].

The decision trees that the target variable takes continuous values called regression trees. Decision trees are often one of the favorites of ML algorithms because of its speed and accuracy. The most common algorithms for decision tree are [91]: classification and regression tree, ID3, C4.5 and C5.0, Chi-squared, M5, and conditional decision trees.

*Application in SBs:* Delgado et al. [92] propose an ML technique based on decision trees to extract the most frequent activities of human behavior and the temporal relationship of those activities in order to produce the human behavior quickly in a smart environment. Viswanathan et al. [93] introduce a prototype distributed data mining system for healthcare environment using C4.5 classification algorithm that can provide the patient monitoring and health services. Decision trees algorithm is a non-parametric algorithm that is easy to interpret and explain. The main disadvantage of this algorithm is that it can easily overfit.

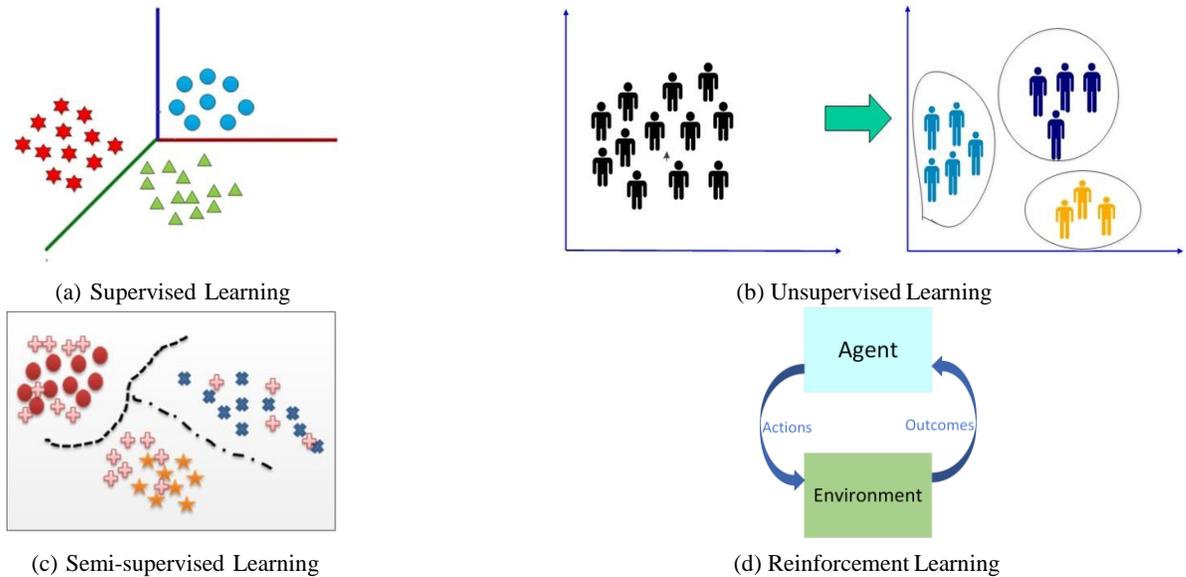

Fig. 7: ML styles.

*c) Bayesian Algorithms:* Bayesian methods utilize Bayes' theorem for classification and regression problems. The most common Bayesian algorithms are [94]: Naive Bayes, Gaussian naive Bayes, Bayesian belief network, Bayesian network.

*Application in SBs:* Parnandi et al. [95] propose an indoor localization approach based on Naive Bayes classification and dynamic time warping, they exploit the embedded sensors of smartphones to determine the building that the user entered and the activities that the user is performing inside the building. Verbert et al. [96] proposed an ML approach based on Bayesian network to diagnosis the fault in HVAC systems. The model has been constructed based expert knowledge concerning conservation laws, component interdependencies, and historical data using virtual sensors. Naive Bayes classifier approaches have been applied with potential results for human activity recognition in [97], [98]. Naive Bayes approach recognizes human activities that identify with the highest probability to the set of sensor readings that were observed.

*d) Support Vector Machine (SVM):* is a supervised ML algorithm which can be applied for both classification and regression problems though mostly used in classification challenges [140]. SVM is one of the most popularly utilized for many statistical learning problems, such as face and object recognition, text classification, spam detections, handwriting analysis etc. [141]. is maximizing the margin that separating between the hyperplane of two classes' closest points. Support vectors are the points lying on the boundaries, and the optimal separating hyperplane is the middle of the margin [142].

*Application in SBs:* Fu et al. [105] proposed an SVM method to predict the system level electricity loads of public buildings that have electricity sub-metering systems. A real-time human tracker system proposed Nguyen et al. [106] using SVM for predicting and recognizing human motion based on the input images from a network of four cameras in the ubiquitous smart homes. Petersen et al. [107] developed an SVM model to predict the times where visitors are present in the home using only the data provided by wireless motion sensors in each room. Fleury et al. [88] presented a study for automatic recognition of daily living activities in a smart home based on SVM. They collected the data from various sensors such as Infra-Red Presence Sensors, door contacts, temperature and hygrometry sensor, and microphones. Das et al. [108] proposed a one-class classification approach for a real-time activity error detection in smart homes using one-class SVM. Zhao et al. [143] proposed an ML approach based on SVM and RNN to detect the occupancy behavior of a building through the temperature and heating source information for the energy efficiency consumption purposes.

*e) Artificial Neural Network Algorithms (ANNs):* ANN models are inspired by the process of biological neural networks. ANN models are commonly utilized for regression and classification problems. The common ANN algorithms are [94]: Perceptron, Back-Propagation, Hopfield Network, and Radial Basis Function Network (RBFN).

ANNs provide a number of advantages including it requires less statistical training, it has the capacity to detect complex nonlinear relationships between the predictor and response variables, as well as the ability to detect all possible relationships between predictor variables [104]. On the other hand, disadvantages include its "black box" nature, heavy computational burden and proneness to overfitting. However, due to the inherent features of neural networks, it has the following main limitations: challenge in training with no local optima, its accommodation to modifications in the behavior, the validation process of the results, and the complexity of network performance interpretation.

*Application in SBs:* Badlani and Bhanot [99] developed a smart home system for energy efficiency applying pattern recognition based on ANNs, the system incorporates an RNN to capture human behavior patterns and an ANN for security applications in smart homes. Other researchers have applied

TABLE V: COMPARISON OF ML TECHNIQUES

| Category | Type | Algorithms | Pros | Cons | Applicability in SBs | Cited |
|---|---|---|---|---|---|---|
| Supervised Learning | Classification | Neural networks | Requires little statistical training; Can detect complex non-linear relationships | computational burden; Prone to Overfitting; Picking the correct topology is difficult; Training can take a long time and a lot of data; | Used for classification, control and automated home appliances, next step/action prediction. | [99][100] [101][102] [103][104] |
| | | SVM | Can avoid overfitting using the regularization; expert knowledge using appropriate kernels | Computationally expensive; Slow; Choice of kernel models and parameters sensitive to overfitting | Classification and regression problems in SBs such as activity recognitions, human tracking, energy efficiency services | [105][106] [107][88] [108] |
| | | Bayesian networks | Very simple representation does not allow for rich hypotheses | You should train a large training set to use it well. | Energy management system and human activity recognition. | [109][96] [97][98] |
| | | Decision trees | Non-parametric algorithm that is easy to interpret and explain. | Can easily overfit | Patient monitoring, healthcare services, awareness and notification services. | [92][93] |
| | | Hidden Markov | Flexible generalization of sequence profiles; Can handle variations in record structure | Requires training using annotated data; Many unstructured parameters | Daily living activities recognition classification | [110][111] [112][113] |
| | | Deep Learning | Enables learning of features rather than hand tuning; Reduce the need for feature engineering | Requires a very large amount of labeled data, computationally really expensive, and extremely hard to tune. | modeling occupant's behavior, and in human voice recognition and monitoring systems; Context-aware SB services. | [114][115] [116][117] [118][119] |
| | Regression | Orthogonal matching pursuit | Fast | Can go seriously wrong if there are severe outliers or influential cases | For regression problem such as energy efficiency services in SBs. | [120] |
| | | clustered-based | Straightforward to understand and explain, and can be regularized to avoid overfitting. | It is not flexible enough to capture complex patterns | Gesture recognition. | [121] |
| | Ensemble methods | N/A | Increased model accuracy through averaging as the number of models increases. | Difficulties in interpreting decisions; Large computational requirements. | Human activity recognition and Energy efficiency services. | [122][123] [124] |
| | Time series | N/A | Can model temporal relationships; Applicable to settings where traditional between-subject designs are impossible or difficult to implement | Model identification is difficult; Traditional measures may be inappropriate for TS designs; Generalizability cannot be inferred from a single study. | Occupant comfort services and energy efficiency services in SBs. | [125][126] [127] |
| Unsupervised learning | Clustering | KNN | Simplicity; Sufficient for basic problems; Robust to noisy training data. | High computation cost; Lazy learner | Human activity recognition. | [128] |
| | | K-pattern clustering | Simple; Easy to implement and interpret; Fast and computationally efficient | Only locally optimal and sensitive to initial points; Difficult to predict K-Value. | Predict user activities in smart environments. | [86] |
| | | Others | N/A | N/A | N/A | [82][83] [129][130] |
| Semi-Supervised learning | N/A | N/A | Overcome the problem of supervised learning having not enough labeled data. | false labeling problems and incapable of utilizing out-of-domain samples. | Provide context aware services such as health monitoring and elderly care services. | [131][132] [133][134] [135] |
| Reinforcement learning | N/A | N/A | Uses "deeper" knowledge about domain | Must have (or learn) a model of environment; must know where actions lead in order to evaluate actions | Lighting control services and learning the occupants, preferences of music and lighting services. | [136][137] [138][139] |

ANNs to present context-aware services. Campo et al. [100] developed a system that calculates the probability of occupation for each section of the building and compares the probability with the current situation systematically. See [101] for a survey paper focusing on the role of ANNs for smart home services. Ermes et al. [102] proposed a hybrid classifier approach using a tree structure comprising a priori knowledge and ANN to recognize the activities such as rowing, biking, playing football, walking, running, sitting, or hiking. Ciabattoni et al. [103] proposed a home energy management system design using the neural network algorithm to predict the power production of the photovoltaic plant and the home consumptions during the given time.

*f) Deep Learning Algorithms:* Deep learning methods represent an evolved form of ANNs in which a deep architecture (many layers comprising multiple linear and non-linear transformations [144]) is used. One of the promises of DL is replacing the manually selected features with efficient unsupervised or semi-supervised feature learning and hierarchical feature extraction algorithms. The most common DL algorithms are [145]: Convolutional Neural Network (CNN), Recurrent Neural Network (RNN), Deep Boltzmann Machine (DBM), Deep Belief Network (DBN), and Stacked Auto-Encoders. Deep learning has been used successfully in varieties of big data analytics applications, particularly natural language processing (NLP) applications, medical diagnosis, stock market trading, network security, and image identification.

Deep learning is now ubiquitously used in major businesses and companies. Microsoft research on a deep learning system presented real-time speech translation system between Mandarin Chinese and English languages [146]. Apple's Siri uses a deep learning trained model, and the voice recognition in the Google Android phone also uses a deep learning trained model [147]. DL utilizes a number of techniques such as dropout and convolutions that enables the models to learn efficiently from high-dimensional data. However, DL requires much more data to train compared to other algorithms because of the magnitudes of parameters for estimation required by the models.

*Application in SBs:* Choi et al. [114] propose two prediction algorithms deep belief network and restricted Boltzmann machines based on the DL framework for predicting different human activities in a building. They also presented a hybrid model which combines for predicting human behavior. The paper [115] proposes a generic deep learning framework based on convolutional and RNNs for human activity recognition that is suitable for multimodal wearable sensors, such as accelerometers, gyroscopes or magnetic field sensors. Alsheikh et al. [116] proposed a hybrid approach of DL and hidden Markov model for human activity recognition using triaxial accelerometers. Baccouche et al. [117] propose a two-steps neural-based deep model to classify human activities, the first step of the model is automatically learned spatiotemporal features based on Convolutional Neural Networks. Then the second step of the model uses an RNN to classify the entire sequence of the learned features for each time-step. In [118], they propose an acceleration-based human activity recognition method using Convolution Neural Network. In [119] a deep convolutional neural network as the automatic feature extractor and classifier for recognizing human activities is proposed using the accelerometer and gyroscope on a smartphone. Hammerla et al. [148] explore the performance of deep, convolutional, and recurrent approaches of deep learning for human activity recognition using wearable sensors. For the sake of measuring the performance, the authors used three representative datasets that comprise motion data collected from wearable sensors.

*g) Hidden Markov models (HMM):* An HMM is a doubly stochastic process with a hidden underlying stochastic process that can be observed through the sequence of observed symbols emitted by another stochastic process.

*Application in SBs:* Wu et al. [113] proposed an improved HMM to predict user behaviors in order to provide services for people with disabilities. They developed a temporal state transition matrix to be utilized instead of the fixed state transition matrix. Lv and Nevatia [112] used hidden Markov models for both automatic recognition and segmentation of 3-D human activities to allow real-time evaluation and feedback for physical rehabilitation. Cheng et al. [110] proposed an inference engine based on the HMM that provides a comprehensive activity of daily living recognition capability. They integrated both Viterbi and BaumWelch algorithms to enhance the accuracy and learning capability. Chahuara et al. [111] proposed sequence-based models for online recognition of daily living activities in an SB environment. They presented three of sequence-based models: HMM, conditional random fields, and a sequential Markov logic network.

*h) Time Series Analysis:* A time series is a collection of temporal instances; time series data set usually have the following characteristics include the high dimensionality, large number of instances, and updating continuously [149]. One of the important purposes for time series representation is to reduce the dimension, and it divides into three categories: model-based representation, non-data-adaptive representation, and data-adaptive representation [150] [151].

*Application in SBs:* Survadevara et al. [125] proposed a wellness model using seasonal autoregression integration moving average time series with sleeping activity scenario in a smart home environment to forecast the elderly sleeping tendency. Zhou et al. [126] proposed a time series analysis framework to explore relationships among non-stationary time series in the case of data sensors in SBs. Jakkula and Cook [127] propose a time series based framework to determine temporal rules from observed physical and instrumental activities of occupants in a smart home.

*i) Regression:* The aim in regression problems is to estimate a real-valued target function. It is related to representing the relationship between variables that are repeatedly processed utilizing a measure of error in the predictions made by the model [152]. The most common regression algorithms are [153]: linear regression, logistic regression, stepwise regression, and ordinary least squares regression.

*Application in SBs:* Chen et al. [120] used the regression technique of orthogonal matching pursuit algorithm to identify the physical and environmental parameters that providing the energy efficiency in an SB. Bouchard et al. [121] presented a gesture recognition system using linear regression combined with the correlation coefficient to recognize the gesture direction and estimate the segmentation of continuing gestures of daily usage activities in a smart environment.

*j) Ensemble methods:* A combination of multiple classifiers often referred to as a classifier ensemble, group of classification models that are trained separately and the predictions of those models are then combined in a way to produce the overall prediction [154]. The most popular ensemble learning based classification techniques are [155]: random forest, boosting, gradient boosting machines, AdaBoost, bagging, and blending.

*Application in SBs:* Jurek et al. [122] proposed a cluster-based ensemble approach solution for activity recognition within the application domain of smart homes. With this approach, activities are modeled as cluster collections built on different subsets of features. Fatima et al. [123] proposed an ensemble classifier method for activity recognition in smart homes using genetic algorithm optimization to merge the prediction output of multiple classifiers that make up the ensemble. They used the ANN, HMM, conditional random field, and SVM [13] as base classifiers for activity recognition. Guan and Ploetz [124] proposed a deep LSTM ensemble method for activity recognition using wearables: more specifically, the authors developed modified training procedures for LSTM networks and proposed the combination of sets of diverse LSTM learners into classifier collectives.

*2) Unsupervised Learning:* Unsupervised Learning refers to developing algorithms that use data with no labels to analyze the behavior or the system being investigated [156]. Thus, the algorithm does not know about the truth of the outcome. In other words, the unsupervised learning algorithm classifies the sample sets to different clusters by investigating the similarity between the input samples. Clustering is done using different parameters taken from the data which enable us to identify correlations which are not so obvious. The inferring structures existing within the input data is used to prepare the model to prepare and extract general rules of the model. A mathematical process might be used to systematically reduce redundancy, or organize data by similarity [129].

The unsupervised approach has been applied to recognize various activities in smart buildings when it is challenging to have labels for input data [130]. Common unsupervised learning problems are clustering, dimensionality reduction, and association rule learning. There are a variety of commonly used unsupervised learning algorithms, some of those algorithms are based on supervised-learning algorithms: the Apriori algorithm and k-Means. In unsupervised learning, usually there is no a measure for the output; we recognize only the features and the target is to define the patterns and relationships among a set of input measures [80].

The major disadvantage of unsupervised learning is the absence of direction for the learning algorithm, hence, there might not be any useful detected knowledge in the selected set of attributes for the training. Clustering is a method of unsupervised learning that involves detecting patterns in the data by placing each data element into a group of K-clusters, where each group holds data elements most similar to each other [157]. Unsupervised learning problems can be categorized into clustering and association problems, which are described next.

*a) Clustering:* A clustering problem explores the internal groupings in the input data, such as grouping customers by their purchasing habits. Clustering techniques are usually organized by modeling strategies such as centroid-based and hierarchical. All methods are concerned with handling the internal structures in the input data to properly organize the data into groups of maximum commonality [158]. The quality of the clustering result is evaluated depends on the type of application that utilizes a clustering algorithm. For example, the sum of squared errors is generally utilized for data clustering while the peak-signal-to-noise ratio is used for image clustering [21]. The most common clustering algorithms are [153]: k-Means, k-Medians, expectation maximization, and hierarchical clustering.

*Application in SBs:* Fahad et al. [128] propose an activity recognition approach that combines the classification with the clustering, in their approach the activity instances are clustered using Lloyd's clustering algorithm. Then, they apply evidence theoretic K-Nearest neighbors learning method that combines KNN with the Dampster Shafer theory of evidence. The paper [86] proposes a hybrid approach to recognize and predict user activities in a smart environment. They use the K-pattern clustering algorithm to classify so varied and complex user activities, and ANN to recognize and predict users' activities inside their personal rooms. Lapalu et al. [82] used an unsupervised learning approach to address the issues of daily living activities' learning in smart home. They utilize the Flocking algorithm for clustering analysis of a use case in cognitive assistance service that assists the people suffering from some type of dementia such as Alzheimer's disease. Aicha et al. [83] present an unsupervised learning approach for detecting abnormal visits of an elderly in a smart home environment based on a Markov modulated Poisson process model. The model combines multiple data streams, such as in the front-door sensor transitions and the general sensor transitions. The other cases of social communication services, Cook et al. [129] applied an unsupervised learning approach to detect social interaction and monitor activity daily living in a smart space, their approach can adapt and update automatically to reflect the changes in discovered patterns from implicit and explicit identified feedbacks of the occupant. Rashidi et al. [130] introduce an unsupervised method that identifies and tracks the normal activities that commonly occur in an individual's routine in a smart environment. The activity discovery method of the system is produced to cluster the sequences based on the simple k-means algorithm. Fiorini et al. [159] proposed an unsupervised ML approach to identify the behavioral patterns of the occupants using unannotated data collected from low-level sensors in an SB. Their approach involves processing and analyzing collected data related to the daily living activities of 17 older adults living in a community-based home supplied with a variety of sensors. They extract activity information from collected data at different times of the day.

*b) Association:* The association rule learning problem is utilized to identify the rules that define large portions of input data, such as people that buy *X* item also tend to buy *Y* item. Association analysis is performed on rules discovered by analyzing input data for frequent if/then statement and using the criteria of *support* and *confidence* to discover relationships between unrelated data in a relational database or another information repository. Here "support" indicates how frequently the items appear in the database while "confidence" indicates the number of times the if/then statements have been found to be true. Many algorithms for generating association rules have been proposed. Apriori algorithm is the most well-known association algorithm [160].

*Application in SBs:* Aztiria et al. [161] proposed system that learns the frequent patterns of human behavior using association, workflow mining, clustering, and classification techniques. The core part of the system is the learning layer which is made up of two modules: the language module, which provides a standard conceptualization of the patterns; and the algorithm module, which discovers the patterns. Kang et al. [162] proposed a service scenario generation scheme for interpreting association rules extracted from the states of all devices in SB environments. Typically, These states are collected periodically at a specific time interval from the devices. Nazerfard et al. [163] propose a framework to discover the temporal features of the activities, including the temporal sequencing of activities and their start time and duration using the temporal association rule techniques in a smart home.

*3) Semi-Supervised Learning:* Semi-Supervised learning lies between supervised and unsupervised methods. Input data is a composite of labeled and unlabeled samples. These hybrid algorithms aim to inherit the strengths of the main categories while mitigating their weaknesses. The model learns the patterns present in the data and also make predictions.

Example problems are classification and regression [164].

There are some common semi-supervised learning models, including generative models, heuristic approaches, semi-supervised SVM, graph-based methods, self-training, help-training, mixture models, co-training and multi-view learning [94].

*Application in SBs:* Cook [131] combined fully-supervised and semi-supervised learning to recognize and follow activities that support health monitoring and assistance context-aware services for people experiencing difficulties living individually at smart homes. Liu et al. [132] proposed a vision based semi-supervised learning approach for fall detection and recognizing other activity daily living in smart environments to overcome the labeling challenges of human activities by systematic interpreting the activities with the highest confidence. Fahmi et al. [133] proposed a semi-supervised fall detection approach in which a supervised algorithm utilizing decision trees in the training process and then profiles are used to implement a semi-supervised algorithm based on multiple thresholds. Radu et al. [134] present semi-supervised ML method using only the low power sensors on a smartphone to consider the problem of determining whether a user is indoors or outdoors. Guan et al. [135] propose a semi-supervised learning algorithm for activity recognition named En-Co-training to make use of the available unlabeled samples to enhance the performance of activity learning with a limited number of labeled samples. The proposed algorithm extends the co-training paradigm by using an ensemble method.

*4) Reinforcement Learning:* Reinforcement learning is a learning approach to control a system in order to maximize performance measure that represents a long-term objective [165]. Reinforcement learning, an area of ML inspired by behaviorist psychology, is concerned with the way that software agents have to take actions in an environment in order to maximize the concept of cumulative reward. RL algorithms learn control policies, particularly when there is no a priori knowledge and there is a massive amount of training data. However, RL algorithms suffer from some drawback such as the high computational cost required to find the optimal solution, such that all states need to be visited to choose the optimal one. The well-known approaches of RL are Brute force, Monte Carlo methods, Temporal difference methods, and Value function [166]. Q-learning [167] is a model-free reinforcement learning approach based on learning the required utility given a state decision.

*Application in SBs:* Mozer [136] applied Q-learning for lighting regulation to predict the time of turning the lights ON/OFF in a building. This prediction model can be utilized to schedule the lights' activations in a building for efficient energy consumption proposes. Li and Jayaweera [137] proposed a Q-learning based approximate dynamic programming algorithm to provide a more efficient, flexible and adaptive method. This approach can enable customers to make an optimal on-line decision making in SB environment to maximize the profits based on both local fully observable and the estimated hidden information of the building. Khalili and Aghajan [138] proposed a temporal differential class of RL method for autonomous learning of a user's preference of music and lighting service settings in presence of different states of the user in SB environment. The preferences are learned by the model by using the explicit or implicit feedback from users when they react to the provided service. Xu et al. [139] give a survey of developments in RL algorithms with function approximation. They evaluated and compared different RL algorithms using several benchmark learning, prediction, and learning control tasks.

*B. ML Tasks for SBs*

In this section, we will describe the major ML tasks that are relevant for SB. The reader is referred to Figure 8 for a general depiction of ML tasks in SBs and the steps taken to implement ML in an SB environment.

*1) Data Collection and Acquisition:* A variety of data collection approaches are used, each of which has different deals in terms of capabilities, energy efficiency, and connectivity. Sensors and similar objects in SBs produce raw data simultaneously in an automated way and such devices may store the data for a specific period of time or report it to controlled components [168]. Data can be collected at gateways; the collected data is then filtered and processed, fused into compact forms for efficient transmission. A variety of communication technologies such as Zigbee, Wi-Fi, and cellular are utilized to transfer data to collection points.

Data collected from a global-scale deployment of smart things defines the basis for decision making and providing services. It is possible that the decisions are unreliable when the quality of utilized data is poor [169].

Zhao et al. [170] propose a data acquisition and transmission system which could be used for monitoring systems to collect energy consumption data (e.g., electricity, water, gas, heating, etc.) from terminal meters which are installed in buildings. The system stores the data periodically after analyzing and processing it and finally transmits the data to servers through the Ethernet. Rowley et al. [171] propose the data acquisition and modeling approaches that can support the delivery of building energy infrastructure in both new building and renovated real-world contexts. Such methods provide a means to achieve short, medium and long-term forecasting of possible scenario pathways to multi-objective sustainable outcomes.

*CLEEN MMEA* [172] platform that collects, processes, and manages the data and initiates contextual knowledge extraction. The purpose is to establish an online marketplace to collect data and provide services for different companies. The interfaces are made public so that any company can easily join the network to buy or sell services. The analysis results can be given to an energy services company in order to allow offering the service to the owners.

A typical example of open access data collection system is *e3Portal* [173] developed by VTT in collaboration with Finnish municipalities. e3Portal offers information and tools when planning savings measures and energy retrofitting in municipal buildings. It also involves frequently updated data regarding energy and water consumption in thousands of public buildings like schools, kindergartens, offices, hospitals, other health care facilities, etc. Decision makers, designers, operation and maintenance personnel, as well as buildings users, can utilize it.

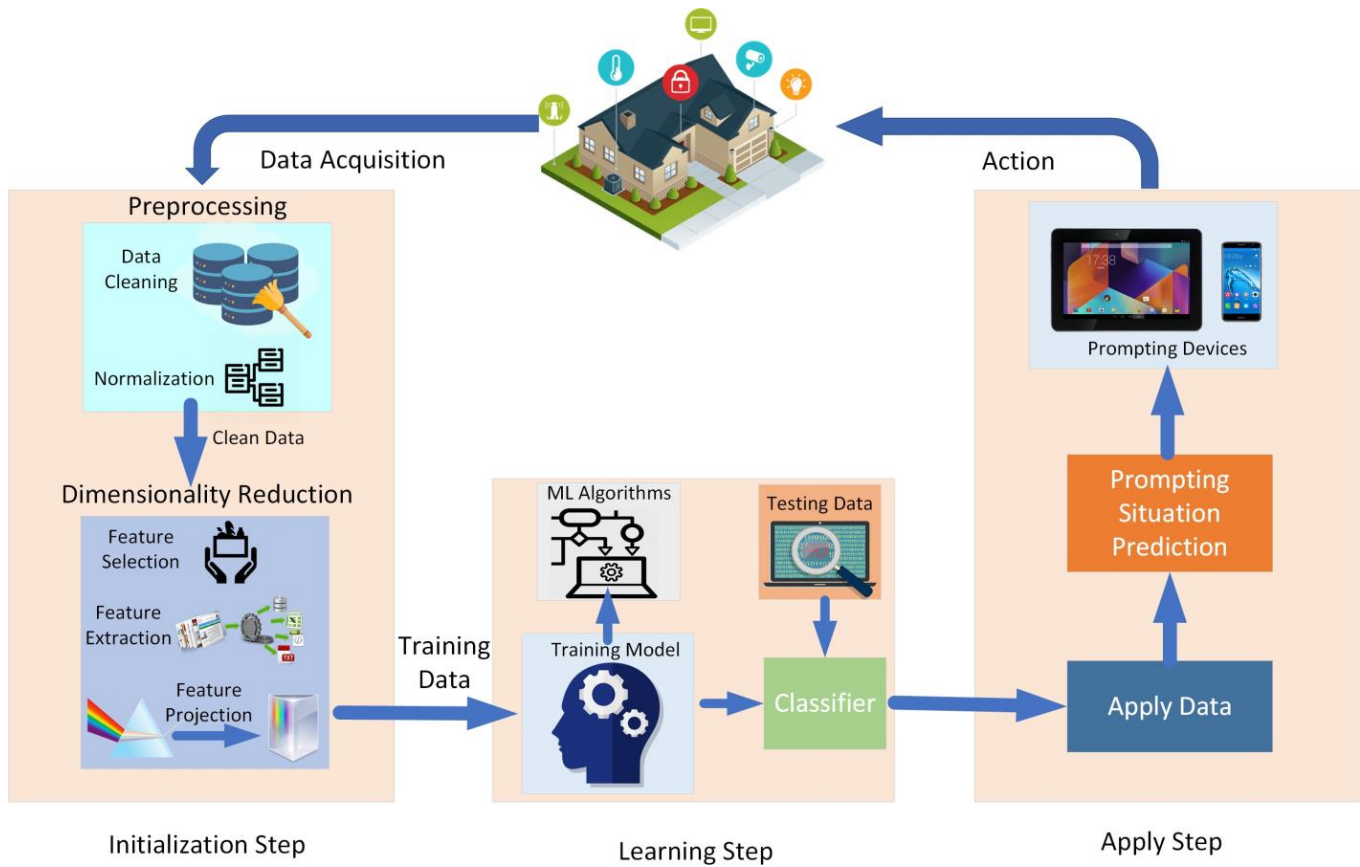

Fig. 8: ML tasks in SB environment.

There are projects that provide publicly available SB datasets for researchers to conduct further studies; A list of "Home Datasets" [174] includes the datasets collected by projects from UC Berkeley, MIT, Washington State University, University of Amsterdam, University of Edinburgh, and the University of Tokyo. The *WARD* [175] project supported by NSF TRUST Center at UC Berkeley provides a benchmark dataset for human action recognition using a wearable motion sensor network. The dataset was collected from 13 repetitive actions by 13 male and 7 female participants between the ages of 19 and 75. An MIT project [176] collected daily live activities dataset from two single-person apartments within a period of two weeks. Eighty-four sensors to record opening-closing events were attached to different appliances and devices such as drawers, refrigerators, containers, etc. Baos et al. [177] introduced an open benchmark dataset collected from various inertial sensors attached to different parts of the body. They considered 33 fitness activities, recorded using 9 inertial sensor units from 17 participants. The *CASAS* project [178] at Washington State University provides a publicly-available dataset for a three-bedroom apartment with one bathroom, a kitchen, and a living room. Different types of motion and digital sensors are installed to support temperature readings, in addition, the analog sensors are installed to support readings for hot water, cold water, and stove burner use [179]. The *PlaceLab* project [180] of MIT provides a dataset collected from a one-bedroom apartment with more than 900 sensors, including those coming from motion, switch and RFID sensors. That is being used to monitor activity in the environment in the context of a smart home [181]. A collection of smart meters data from five houses in the UK [182] That consists of 400 million instances. The active power is formed by different appliances and the whole-house power demand every 6 seconds.

The major challenges that arise for data collection are scalability, privacy, security, and heterogeneity of resources [183]. Automated sensor data collection process collects a large amount of data that overwhelms the collection and analysis centers in comparison to the data collected from other sources such as IoT devices and social media. This leads to a huge number of small synchronous write operations to the database storage system, consequently, resulting in serious performance bottlenecks to the storage system design [184]. Because of the extensive use of RFID technology, privacy issues arise in data collection; for example, the RFID tags carried by a person may become a unique identifier for that person. Also, other security concerns appear, for example, the radio signals of RFID technology are easily jammed. Hence, that can disrupt the data collection process [185].

The heterogeneity of data that is being collected from different resources is another major challenge, such that the data are usually very noisy, large-scale, and distributed. This makes it very difficult to use the collected data effectively without a clear description of existing data processing techniques [184].

*2) Data Preprocessing:* A large amount of data are generated by sensors in SBs; this data comes from various sources with diverse formats and structures. Usually, this data is not ready for analysis as it might be incomplete or redundant due to low battery power, poor calibration, exposure to various malicious elements and interference. Therefore, raw data typically needs to be preprocessed to deal with missing data, discard noisy and redundant data and integrate data from various sources into an integrated schema before being committed to storage. This preprocessing is called data cleaning. The quality of data can be improved substantially by applying some cleaning techniques to the data before it arrives its end user [168][186]. Data cleaning is one of the significant tasks in the data processing phase. Data cleaning is not a new process particular for the IoT data processing. It has already been applied as a process for database management systems. Presenting a data cleaning method would further aid the applications to focus on their core logic without worrying about data reliability post-processing overheads [184].

There are many different techniques that have been utilized to deal with the problem of cleaning noisy data streams such as Kalman filters [187], statistical models [186] and outlier detection models [184]. One of the major challenges with data cleaning techniques in the SBs is the heterogeneity of data collected from different sources particularly WSN- and RFID-enabled data streams. The utilized data cleaning techniques should be able to deal with several different variables of interest to satisfy IoT applications' requirements, for example, setting home temperature based on observed outer temperature, user habits, energy management, etc. [169] Any type of failures such as a failed sensor, network issues, camera failure, or database crashes in the process of collecting data would invalidate the data. Consequently, this type of impediment will dramatically increase the time required to collect data [179].

*3) Dimensionality Reduction:* There are huge volumes of raw data that are captured from heterogeneous and ubiquitous of sensors used in SBs. Most of the data collected from those sensors are redundant and they need to be brought down to a smaller number of features by applying dimensionality reduction techniques without losing significant information [188]. The main idea from the dimensionality reduction strategy is to find a new coordinate system in which the input data can be represented with much fewer features without losing significant information. The dimensionality reduction can be made in two different ways: by extracting of the features that represent the significant data characteristics (this technique is called feature extraction), or by only selecting the most relevant features from the original dataset, this method is called feature selection [189] [190].

Like clustering methods, the dimensionality reduction approach explores and exploits the internal structure of the data, but in this case in an unsupervised manner using less information. Most of these techniques can be utilized in classification and regression problems. Examples of some salient algorithms are [153]: Principal Component Analysis (PCA), Principal Component Regression (PCR), and Linear Discriminant Analysis (LDA). Chen et al. [191] propose a framework using the classification information of local geometry of data to reduce the dimensionality of a dataset on human activity recognition from wearable, object, and ambient sensors.

*a) Feature Extraction:* The main components of the original data are the features. After extracting the features from the raw dataset, such features contain important information that is used by the learning algorithms for the activities discrimination. The most common methods of feature extraction work in time, frequency, and discrete domains [192]. Among time domain method, mean and standard deviation are the key approaches for almost all sensor types. While the frequency domain method focuses on the periodic structure of the collected data. Wavelet Transformation and Fourier Transform are the most common approaches. And discrete domain methods such as Euclidean-based distances, dynamic time warping, and Levenshtein edit distance are key approaches implemented in several applications such to string similarity, classifying human activities and modeling human behavioral patterns [16], [193].

*b) Feature Selection:* The main role of feature selection is to discriminate the most related subset of features within a high dimensional vector of features, so that reduces the load of noise and computational expense on the learning models. In order to map the high dimensional vector of features into a lower dimensional vector, there are several common algorithms used such as Linear Discriminant Analysis (LDA), Principal Component Analysis (PCA), and Independent Component Analysis (ICA) [194].

Hausmann and Ziekow [195] proposed an approach for automatically adapting the feature selection for SBs application ML models from the time-series data based on wrapper methods and genetic optimization. Fahad et al. [196] propose an activity recognition approach for overlapping activities using K-Nearest neighbors approach that distinguishes the most important features from the collected information obtained from deployed sensors in multiple locations and objects. Fang et al. [197] determine that the different feature sets generate different levels of accuracy for recognizing human activities, and selecting inappropriate datasets increases the level of computational complexity and decreases the level of prediction accuracy in smart home environments. The wrapper and filtering are the two main statistical methods of feature selection problem. It is argued that although the wrapper approach may obtain better performances, filters are less resource intensive and faster [198]. In [199], different feature selection methods are utilized for the process of dimensionality reduction of the learning problem to recognize the human activities from observed sensors. The authors show that the performance of the learning models to recognize the human activity has a strong relationship with the utilized features.

*c) Feature projection:* feature projection can be represented as a mapping from the original set of features to an appropriate set that optimizes the learning criterion, such that the feature projection approach allows the process of visualizing and mapping the high-dimensional feature vectors to low dimensional one, in addition, it enables analyzing the distribution of the reduced feature vectors [200]. Consequently, the feature projection approach reduces the pattern recognition's processing time and enables selecting the best-performed classifier for the reduced feature vectors. Hence, it makes real-time implementation possible [198]. Chu et al. [201] proposed a linear supervised feature projection that utilizes the LDA algorithm for EMG pattern recognition that attempted to recognize nine kinds of hand motion.

*C. ML Tools & Platforms for SBs*

There are a variety of existing ML platforms and tools to support the learning process. With the current increasing number of those kinds of toolkits, the task of selecting the right tool for processing big data streaming from various sources can still be difficult. Typically, there is no single toolkit that truly fits and provides solutions for all different problems. Many of the available toolkits might have overlapping uses, and each has advantages and disadvantages. Most of those toolkits might require experiences in the domains of programming languages and system architecture. In addition, usually many people lack a full understanding of the capabilities and how to use those available platforms [202].

The important factors that must be considered when selecting a specific ML tool are scalability, speed, coverage, usability, extensibility, and programming languages support. With respect to the scalability factor, the size and complexity of the data should be considered to determine if a specific toolkit will be fit. The processing platform that the library is running on and the complexity of the algorithm affect the speed factor. Not all the projects prioritize the speed factor; if the models require frequent updates, the speed may be a crucial concern; but not otherwise. Coverage represents the number of ML algorithms implemented in the tool. With the massive amount of data capturing from heterogeneous sources, ML community faces the challenges of how the ML model can efficiently process and learn from the big data.

In general, the available big data tools do not implement all varieties of different classes of ML algorithms, and typically their coverage ranges from a few algorithms to around two dozen. The *usability factor* includes elements such as initial setup processing; continuous maintenance; the available programming languages and user interface available; the amount of documentation, or availability of a knowledgeable user. The *extensibility factor* means that the implementations introduced in the tools can be utilized as building blocks towards new systems. It is necessary to evaluate tools in terms of how well they are able to meet this factor. There are a variety of ML libraries that are available in different programming languages. Depending on the task you are trying to accomplish, certain languages, libraries, and tools can be more effective than others. The following provides a detailed observation of the strengths and weaknesses of the top used deep learning and ML tools. The reader is also referred to Table VI for a concise tabulated summary of the described deep learning and ML tools.

*1) H2O:* H20 [203] is an open-source in-memory, distributed, and scalable ML framework for big-data analysis that supports ML libraries, along with tools for parallel processing, analytics, data preprocessing and evaluation tools. It is produced by the H2O.ai, which launched in 2011 in Silicon Valley. The most notable feature of this product is that it provides numerous tools for deep neural networks. The H2O software APIs can be called from Python, Java, R, and Scala. Users without programming expertise can still utilize this tool via the web-based User Interface. In addition to the processing engine provided by H2O framework, it also allows the users to integrate their models with other available frameworks such as Spark and Storm. Depending on what is suitable for the algorithm, The H2O's engine uses multiple execution methods to process data completely in memory. The general technique used is distributed Fork/Join, which is reliable and suitable for massively parallel tasks.

The H2O software can be run on different operating systems such as Microsoft Windows, Mac OS X, and Linux (e.g. Ubuntu 12.04; CentOS), It also runs on Apache Hadoop Distributed File System (HDFS) and Spark systems for big-data analysis. In addition, it can operate on various cloud computing environments such as Amazon EC2, Google Compute Engine, and Microsoft Azure. As of July 2016, the algorithms supported in H2O cover the tasks classification, clustering, generalized linear models, statistical analysis, ensembles, optimization tools, data preprocessing options and deep neural networks.

*2) MLlib (Spark):* MLlib [204] is Apache Spark's ML library. MLlib aims to provide scalable and easy to use ML methods. It includes common ML algorithms for classification, regression, clustering, dimensionality reduction, as well as lower-level optimization primitives and higher-level pipeline APIs. The classification techniques of SVM, random forest, logistic regression, Naïve Bayes, decision trees, and gradient-boosted trees are supported whereas for clustering, k-means, Gaussian mixture, and power iteration clustering are supported. MLLib supports implementations for linear regression and isotonic regression, and incorporates a collaborative filtering algorithm using alternating least squares. PCA is supported for dimensionality reduction. MLlib includes APIs for development in Scala, Java, Python, and SparkR. Generally, MLlib depends on Spark's iterative batch and streaming approaches, as well as its use of in-memory computation.

*3) TensorFlow:* Tensorflow [205] is an open source software library for numerical computation and deep ML in a variety of perceptual and language understanding tasks utilizing data flow graphs. TensorFlow was originally developed by the Google Brain team and was released in November 2015 under an Apache 2.0 open source license. TensorFlow has tools that support deep learning, reinforcement learning, and other algorithms. TensorFlow implements data flow graphs, where "tensors" are batches of data that can be processed by a set of algorithms defined by a graph.

The movements of the data through the system are called "flows"—hence, the name. TensorFlow can run on multiple CPUs and GPUs. It can run on Linux, Mac OS X desktop, and server systems, and Windows support on roadmap, as well as on Android and Apple's iOS mobile computing platforms. TensorFlow is written with a Python API over a C/C++ engine that makes it run fast. TensorFlow utilizes a symbolic graph of vector operations method, in order to easily define a new network. However, TensorFlow has a weakness that is related to modeling flexibility. Such that each computational flow has to be constructed as a static graph. That makes some computations such as beam search difficult.

*4) Torch:* Torch [206] is an open source ML computing framework that supports a variety of ML algorithms. Torch was originally developed at NYU. It is efficient and easy to use, thanks to a script language based on the Lua programming language and a C/CUDA implementation, Torch was intended to be portable, fast, extensible, and easy to use in development. Some version of Torch is employed by large companies such as Google DeepMind, the Facebook AI Research Group, IBM, Yandex, and the Idiap Research Institute. In addition, it has been extended to run on Android and iOS platforms. A variety of community-contributed packages for Torch, giving it a versatile range of support and functionality. It provides various deep learning algorithms that support computer vision; signal, image, video, and audio processing; parallel processing and networking [207].

*5) Deeplearning4j:* Deeplearning4j [221] is an open source distributed DL library, primarily developed by Adam Gibson from an ML group in San Francisco. Deeplearning4j is written for Java and JVM as well as to support a variety of DL algorithms such as restricted Boltzmann machine, deep belief networks, convolutional networks, recurrent neural networks, deep autoencoder, stacked denoising autoencoder, and recursive neural tensor network. All these algorithms can be integrated with Hadoop and Spark for distributed parallel processing. Deeplearning4j relies on Java programming language, in addition, it is compatible with Clojure and includes a Scala API. Deeplearning4j is designed to be utilized in business environments, rather than as a research tool. It is applied in a variety of applications such as fraud detection, anomaly detection, recommender systems, and image recognition.

*6) Massive Online Analysis (MOA):* MOA [222] is one of the common open source frameworks for data stream mining and possessing. MOA is written in Java related to the WEKA project that developed at the University of Waikato, New Zealand. It includes a set of learners and stream generators that can be used from the GUI, the command-line, and the Java API. MOA supports a variety of ML algorithms for classification, regression, clustering, outlier detection, as well as some tools for evaluation [223].

*7) Caffe:* Caffe [224] is a DL framework, it is primarily developed with the consideration of expression, speed, and modularity. It utilizes the machine-vision library for fast convolutional networks from Matlab, which has been ported to C and C++. It is developed by the Berkeley vision and learning center and by the community contributors. In Caffe, multimedia scientists and practitioners have an organized and state-of-the-art toolkit for DL algorithms. Caffe was originally developed for machine-vision, it has been utilized and improved by users in other fields such as robotics, neuroscience, speech recognition, and astronomy. In addition, it supports Python and MATLAB code bindings. Caffe offers image classification with state of the art CNN algorithm. Caffe is mainly utilized as a source of pre-trained models hosted on its Model Zoo site.

Caffe is useful for performing image analysis using CNNs and regional analysis within images using RCNNs. The performance and processing speed of Cafee make it as one of the most utilized platforms for research experiments and industry deployment. It has the capability to process over 60 million images per day with a single NVIDIA K40 GPU. Caffe has already been applied in many research projects at UC Berkeley and other universities, performing very well in many tasks such as object classification, object detection, and Learning Semantic Features. It provides a complete and well-documented toolkit for training, testing, tuning, and deploying models. Caffe utilizes a large repository of pre-trained neural network models called the Model Zoo, which is suitable for a variety of common image classification tasks [225].

*8) Azure ML:* Microsoft first launched *Azure ML* [226] as a preview in June 2014. Azure ML enables users to create and train models, then convert those models into APIs that can be applied to other services. Users can get up to 10GB of storage per account for model data, although they can also connect their own Azure storage to the service for larger models. programmers can use either the R or Python programming language for developing with Azure services. Users can purchase ML algorithms from Microsoft Azure Marketplace, they can also obtain free algorithms from the community gallery that has been created by Microsoft to share ML algorithms with each other. They share many of predictive analytics of personal assistant in Windows Phone called Cortana. Azure ML also utilizes solutions from Xbox and Bing.

Azure currently supports different features and capabilities such as run Hadoop over Ubuntu Linux on Azure, it also supports hosting Storm for analyzing data streams. In addition, it allows developers to connect .NET and Java libraries to Storm. Azure ML studio supports a variety of modules for training, scoring, and validation processes. Azure ML comes with a large library of algorithms for predictive analytics. The popular families of algorithms are regression, anomaly detection, clustering, and classification.

*D. Real-time Big Data Analytics Tools for SBs*

Several applications need to have real-time data analysis for stream data and waiting for the information to be archived and then analyzed is not practical for these type of applications. Generally, Stream processing is intended to analyze a massive amount of data and act on real-time streaming data utilizing continuous queries such as SQL-type queries to handle streaming data in real-time utilizing scalable, available and fault-tolerant architecture. Essential to stream processing is Streaming Analytics. More and more tools offer the possibility of real-time streaming data. The following presents some of the common and widely used options.

TABLE VI: COMPARISON BETWEEN DEEP LEARNING AND ML TOOLS

| Tool | Creator | OS | Open source? | Written In | Interface | CUDA support? | Algorithms | Release date | Used in |
|---|---|---|---|---|---|---|---|---|---|
| TensorFlow | Google Brain team | Linux, Mac OS X (Windows support on roadmap) | Yes | C++, Python | Python, C/C++ | Yes | deep learning algorithm: RNN, CN, RBM and DBN | November 2015 | [208][209] |
| Theano | Universit de Montral | Cross-platform | Yes | Python | Python | Yes | deep learning algorithm: RNN, CN, RBM and DBN | September 2007 | [210][211][212] |
| H2O | H2O.ai | Linux, Mac OS, Microsoft Windows And Cross-platform incl. Apache HDFS; Amazon EC2, Google Compute Engine, and Microsoft Azure. | Yes | Java, Scala, Python, R | Python, R | No | Algorithms for classification, clustering, generalized linear models, statistical analysis, ensembles, optimization tools, data preprocessing options and deep neural networks. | August 2011 | [213] |
| Deeplearning4j | Various, Original author Adam Gibson | Linux, OSX, Windows, Android, CyanogenMod (Cross-platform) | Yes | Java, Scala, C, CUDA | Java, Scala, Clojure | Yes | Deep learning algorithms including: RBM, DBN, RNN, deep autoencoder | August 2013 | [214] |
| MLlib Spark | Apache Software Foundation, UC Berkeley AMPLab, Databricks | Microsoft Windows, OS X, Linux | Yes | Scala, Java, Python, R | Scala, Java, Python, R | No | classification, regression, clustering, dimensionality reduction, and collaborative filtering | May 2014 | [215] |
| Azure | Dave Cutler from Microsoft | Microsoft Windows, Linux | No | C++ | C++, Java, ASP.NET, PHP, Node.js, Python | Yes | classification, regression, clustering | October 2010 | [216] |
| Torch | Ronan Collobert, Koray Kavukcuoglu, Clement Farabet | Linux, Android, Mac OS X, iOS | Yes | C, Lua | Lua, LuaJIT, C, utility library for C++/OpenCL | Yes | deep algorithms | October, 2002 | [217][218] |
| MOA | University of Waikato | Cross-platform | Yes | Java | GUI, the command-line, and Java | No | ML algorithms (classification, regression, clustering, outlier detection, concept drift detection and recommender systems) | November 2014 | [219] |
| Caffe | Berkeley Vision and Learning Center, community contributors | Ubuntu, OS X, AWS, unofficial Android port, Windows support by Microsoft Research, unofficial Windows port | Yes | C++, Python | C++, command line, Python, MATLAB | Yes | Deep learning algorithms: CN, and RNN | December 2012 | [220] |

*1) Apache Storm:* Storm [227] is an open source distributed real-time data processing framework that provides massively scalable event collection. The initial release was on 17 September 2011, it was created by Nathan Marz and the team at BackType, and is now owned by Twitter. Storm can easily process unlimited streams and with any programming language. It has the capability to process over one million tuples per second per node with a highly scalable, fault-tolerant, and reliable architecture. Storm is written in Java and Clojure. Trident is a high-level abstraction layer for Storm, can be utilized to accomplish state management persistence. Storm is a system of complex event processing. This type of solution allows companies to respond to the arrival of sudden and continuous data (information collected in real-time by sensors, millions of comments generated on social networks such as Twitter, WhatsApp and Facebook, bank transfers etc.). Some of the specific applications of Storm include customer service management in real-time, operational dashboards, data monetization, cybersecurity analytics, and threat detection.

*2) Apache Kafka:* Kafka [228] is a fast, scalable, fault-tolerant and durable open-source message broker project that originally developed by LinkedIn, and subsequently open sourced in early 2011 and released by Apache Software Foundation on 23 October 2012. Kafka is written in Scala. It supports a variety of use case scenarios with a focus on high throughput, reliability, and scalability characteristics. For example, it can message sensor data from heating and cooling equipment in office buildings.

*3) Oracle:* In 2013, Oracle started utilizing Oracle Enterprise Manager that includes Oracle Big Data Appliance to manage all of its big-data technologies. Oracle has also produced multiple low-latency technologies for Oracle Fast Data components includes Oracle Event Processing, Coherence, NoSQL, Business Analytics, and Real-Time Decisions. Oracle Event Processing provides solutions for building applications to filter, correlate and process events in real-time. It supports IoT services by delivering actionable insight on data streaming from a variety of data sources in real-time [229].

Oracle Stream Explorer (OSX) and Oracle R Enterprise (ORE) aim to support equipment monitoring applications for the systems that made of a variety of components through sensors, anomaly detection and failure prediction of such systems. ORE [230] is utilized to handle low-frequency streams in batch mode, while OSX handles the high-frequency streams making real-time predictions and sends the results back to user applications that are communicating with the output channels.

OSX [231] is a middleware platform has the capability to process large amounts of streaming data in real-time for a variety of streaming data applications, from a multitude of sources like sensors, social media, financial feeds, etc. It streamlines real-time data delivery into most popular big data solutions, including Apache Hadoop, Apache HBase, Apache Hive, Apache Flume, and Apache Kafka to facilitate improved insight and timely action. Oracle Real-Time Decisions [232] is a decision management platform with self-learning that determines optimized recommendations and actions with messaging, imagery, products, and services within business processes.

*4) Amazon Kinesis Streams:* Amazon *Kinesis* [233] is a platform for collecting and processing large streams of data on AWS in real-time, AWS launched Kinesis in November of 2013, offering powerful services for loading and analyzing streaming data, in addition, it provides custom streaming data applications for specialized needs. Sometimes Terabytes of data per hour can be generated that need to be collected, stored, and processed continuously from various application services such as web applications, mobile devices, wearables, industrial sensors etc. Typically, Amazon Kinesis Streams application can use the Amazon Kinesis Client Library and reads data from an Amazon Kinesis stream as data records. These applications can run on Amazon EC2 instances.

*5) Apache Spark Streaming:* Apache *Spark* [234] is an open-source platform for real-time data processing, it can implement using four different languages: Scala, the syntax in which the platform is written; Python; R; and Java. Spark Streaming is an extension of core Spark API. It allows building fault-tolerant processing of real-time data streams. Spark Streaming allows the processing of millions of data among the clusters, and Spark SQL which makes it easier to exploit the data through the SQL language. Spark Streaming divides the live data stream into a predefined interval of batches, then handles each batch of data as Resilient Distributed Datasets (RDDs). Then we can apply operations like map, reduce, join, window etc. to process these RDDs. The last results of these operations are then returned in batches.

Spark Streaming can be utilized for a variety of application such as real-time monitoring and analyzing of application server logs. These logs messages are considered time series data. Examples of such type of data are sensor data, weather information, and clickstream data. This data can also be utilized for predicting future states based on historical data. Apache assures a computation speed that performs the operations quicker by 100 times than what is currently offered by Hadoop MapReduce in memory, and 10 times better than in disc. Spark can be executed either in independent cluster mode or in the cloud on different frameworks such as Hadoop, Apache Mesos, and EC2. In addition, Spark can access numerous databases such as HDFS, Cassandra, HBase or S3, Amazon's data warehouse.

*6) Apache Flume:* Flume [242] is a distributed, reliable and open-source log data aggregation framework. Apache Flume is applied in many applications ranging from log data aggregation, to transport massive quantities of event data including network traffic data, social-media-generated data, email messages and pretty much any data source possible into the HDFS.

The architecture of Flume is simple and flexible, it is also robust and fault tolerant with tunable reliability mechanisms for failover and recovery. log manufacturing operations is an example of Flume's application. The a massive log file data can stream through Flume. The log file data can be stored in HDFS and analyzed by utilizing Apache Hive.

*7) Apache SAMOA:* SAMOA [243] is a distributed streaming ML framework that contains programming abstractions for distributed streaming ML algorithms. Its name stands for Scalable Advanced Massive Online Analysis and was originally developed at Yahoo! Labs in Barcelona in 2013 and has been part of the Apache incubator since late 2014. SAMOA is both a platform and a library. It enables the algorithm developer to reuse their code to run on different underlying execution engine. In addition, it supports plug-in modules to port SAMOA to different engines. By utilizing SAMOA, the ML algorithm developer does not need to worry about the complexity of underlying distributed stream processing engines. They can run it locally or utilizing one of stream processing engines, such as Storm, S4, or Samza.

SAMOA provides the ML algorithms for a variety of tasks including classification, regression, clustering, along with boosting, and bagging for ensemble learning. Additionally, it offers a platform for the implementation of these ML algorithms, as well as a framework that enables the user to write their own distributed streaming algorithms. For example, there is CluStream for clustering, as well as Vertical Hoeffding Tree, which uses vertical parallelism on top of the decision tree, or Hoeffding tree for classification. There is also Adaptive Model Rules Regressor, which uses both vertical and horizontal parallelism implementations for regression [244].

A summarized comparison between various real-time data analytics tools is provided in Table VII.

## V. APPLICATIONS OF ML-BASED CONTEXT-AWARE SYSTEMS FOR SBS

The potential uses of ML in an SB environment can be divided into four categories: detection, recognition, prediction, and optimization [79]. We discuss these categories separately next.

In general, *detection* is the extraction of particular information from a larger stream of information. Many detection applications in SBs such as fire detection, leak detection, and anomaly detection [245]. Many different applications have been studied by researchers in activity recognition in SBs; examples include fitness tracking, health monitoring, fall detection. [246].

The goal of *recognition* is to classify an object or an event to a predefined category. It focuses on how to make computer programs perform intelligent and human-like tasks, such as the recognition of an object from an image.

The goal of *prediction* is to determine the temporal relations' model between specific events to predict what will happen in the near future. Prediction can be either for classification or regression problems [247]. Event prediction when the goal is to predict the most probable event or subsequent activity is an example of classification problems, while latency prediction when the output takes on continuous values is an example of the regression problem. The general steps of applying ML processes to predict an event in an SB environment is shown in Figure 9.

TABLE VII: COMPARISON BETWEEN REAL-TIME DATA ANALYTICS TOOLS

| Tool | First released in | Main Owner | Platform | Written in | API languages | Auto-Scaling? | Event Size | Fault Tolerance | Type | Used in |
|---|---|---|---|---|---|---|---|---|---|---|
| Storm | Sept. 2011 | Backtype, Twitter | Cross-platform | Clojure and Java | Any programming language | No | Single | Yes | Distributed stream processing | [235][236] |
| Kafka | Jan. 2011 | LinkedIn, Confluent | Cross-platform | Scala | Java, C++, Node.js | Yes | Single | Yes | Message broker | [237][236][238] |
| Oracle | Jan. 2013 | Oracle | Cross-platform | Java | Java, Node.js, Python, PHP, and Ruby | Yes | NA | Yes | Distributed stream processing | [239] |
| Spark | May 2014 | AMPLab, Databricks | Microsoft Windows, OS X, Linux | Scala, Java, Python, R | Scala, Java, Python, R | Yes | Mini-batch | Yes | Streaming analytics. | [237][215] |
| Amazon Kinesis | Dec. 2013 | AWS | Microsoft Windows, OS X, Linux | C++ | C++, Java, Python, Ruby, Node.js, .NET | Yes | Data blob of 1 MB size | Yes | Real-time streaming data | [240] |
| Flume | Jan. 2012 | Apple, Cloudera | Cross-platform | Java | Java | No | Single | Just with file channel only | Distributed stream processing | [215] |
| SAMOA | May 2013 | Created at Yahoo Labs | Cross-platform | Java | Java | Yes | NA | Yes | Distributed stream processing | [241] |

The goal of *optimization*, on the other hand, is to maximize the long-term profits by making proper decisions in different situations. reinforcement learning can be utilized with these problems. Some optimization problems can be managed as prediction problems such that the profits for different actions are predicted and the action with the highest profit would be selected. Decision making is the most common case of the optimization problem. It takes to consider a variety of variables and solving deals between the profits of different locations of the environment [248].

Smart buildings are becoming increasingly supplied with a variety of sensors that measure different parameters, and data from these sensors is analyzed by ML algorithms and used for a range of services and applications for the activities of the building occupants. SBs go far beyond saving energy and contributing to sustainability goals. The application and services provided by the SBs can be both residential and commercial ranging from e-health, e-marketing, intelligent car parking system, intelligent transportation system, automation, and logistics services.

Figure 10 shows the taxonomy of basic domains of SB services. Lighting service is associated with the well-being of occupants depending on their activities in SBs that have sensors to conserve energy when lights are not needed. The power and electrical system may have onsite renewable energy sources to provide a percentage of power consumption in SBs. HVAC stands for the humidity, ventilation and air conditioning system, intended for the convenience of occupants that have effective interaction with the environment. The water management service is related to increase savings and manage water reclamation for flushing, landscaping and air-cooling systems. Waste management is related to the activities and actions required to collect, separate, transport, together with monitoring and regulation of waste management system in SBs. Parking service is related to minimize the area and volume required for parking cars. I could support car sharing, electric vehicles and a place for bicycles as well. The security is related to managing automated locks, biometric devices as well as video surveillance systems in SBs. The operations control center supports system analytics and decision making for the operations. Visual interfaces provide a dashboard that shows the status of SBs services and human operators to better manage the building resources. These interfaces also allow the occupants to set up their optimal parameters for comfort and productivity improvement on daily activities. Finally, the communications center is related to connecting sensors and actuators in the building as well as the operations control center.

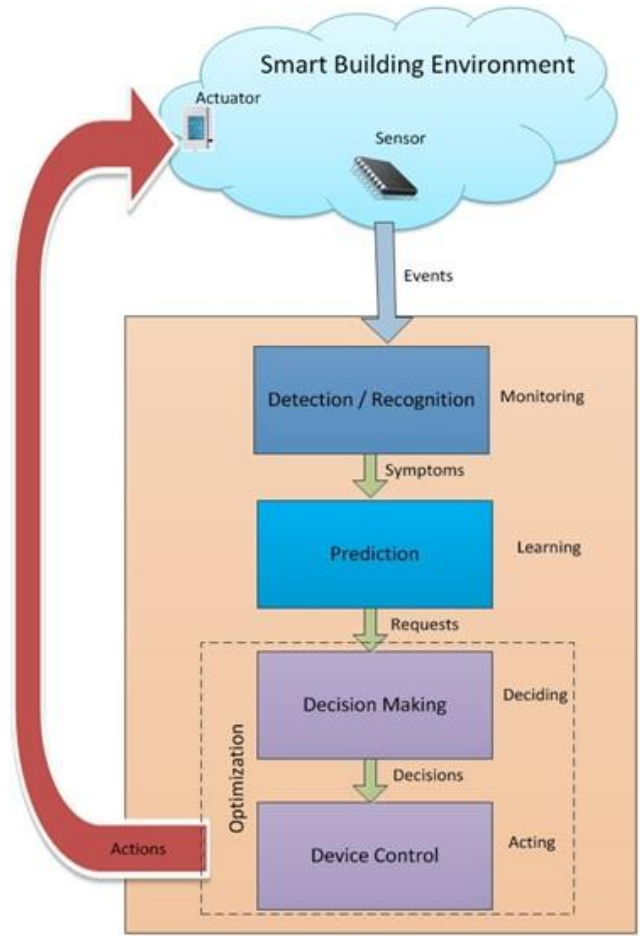

Fig. 9: Steps involved in applying ML models in an SB environment.

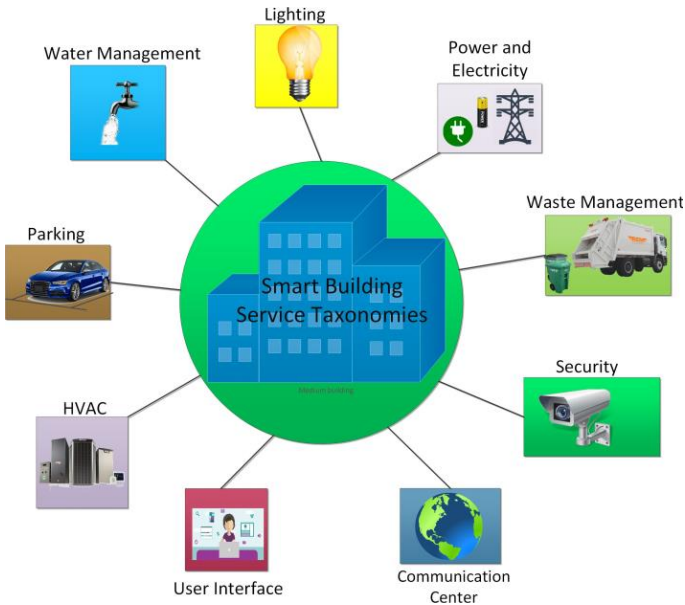

Fig. 10: SB services taxonomy.

Based on our literature survey, we have identified that the application areas of SBs can be elderly care, comfort/entertainment, security/safety, energy management, and other projects. In the rest of the section, we will briefly describe the major domains in providing the following SB services: (1) care of the elderly population; (2) enhancing energy efficiency; (3) enhancing comfort or providing entertainment; (4) enhancing safety and security; and (5) miscellaneous projects.

*a) Elderly Population's Home Care:* SB technology such as sensors, voice activation, GPS, Bluetooth, cellular connectivity via mobile phones, smartphone monitoring apps and sophisticated computers can be especially useful for elderly or disabled individuals who live independently. Elderly persons can take the advantages of such technologies (e.g., monitoring system, emergency system, dangerous kitchen appliance detection, fall detection), to maintain a safe and healthy lifestyle while living independently [56] [249].

Smart technology in the SBs aims to collect real-time information on human daily activity and then learn of their personal patterns. ML techniques have the potential for a very wide array of new innovations in healthcare that will be transformative for both providers and their patients. Whenever a deviation from the norm patterns is detected, SB systems send the alerts to family members and the caregivers in order for them to take urgent response action. By using big data analytics and ML algorithms it is possible to analyze large-scale data contained in electronic medical records—e.g., to learn automatically how physicians treat patients including the drugs they prescribe [250].

Some prominent projects in this space are described next. Chernbumroong et al. [56] proposed an activity recognition and classification approach for detecting daily living activities of the elderly people applying SVM. They used wrist-worn multi-sensors namely accelerometer, temperature sensor and altimeter for detection basic five activities namely feeding, grooming, dressing, mobility, and stairs. And other instrumental activities such as washing dishes, ironing, sweeping and watching TV. Taleb et al. [251] proposed a middleware-level solution that integrates both the sensing and the monitoring services for assisting elders at smart homes environment. The appliances used in the proposed framework include RFID readers that cover of the whole building, sound sensors, video cameras, smart door lock, microphone and speakers for interaction with the system. CAALYX [252] is a European Commission-funded project that supports older people's autonomy and self-confidence. The service is formed of three distinct subsystems including elderly monitoring subsystem, home monitoring subsystem and the caretaker's monitoring subsystem. The system delivers a high priority message to an emergency service including the geographic position and clinical condition of the elder user. EasyLine+ [253] project funded by the European Commission to support elderly people with or without disabilities in carrying out a longer independent life at home. The system uses a neural network, assistive software, and a variety of sensors such as illumination sensor, temperature sensor, door sensors, and RFID giving the capacity of controlling the white goods. Hossain et al. [254] proposed a cloud-based cyber-physical multi-sensory smart home framework for elderly people that supports gesture-based appliance control. Suryadevara et al. [255] proposed a model for generating sensor activity pattern and predicting the behavior of an elderly person using household appliances.

*b) Energy Efficiency:* When temperatures rise or fall in various zones of your home, heaters, air conditioners, fans, and other devices will turn on or off (or increase or decrease in speed or temperature). In order to perform an efficient energy consumption of the supply systems, a significant step that is necessary by analyzing the way that current energy consuming system is using in buildings [256]. In the last decade, analysis of the energy efficiency in the smart spaces has received increasing attention. Various approaches for energy efficiency have been proposed utilizing predictive modeling based on profile, climate data, and building characteristics [32] [257]. For instance, lights throughout your home might turn on and off depending on the time of day.

In the past, various attempts have been made to improve energy efficiency in the SBs through the use of smart metering and sensor networks at the residential level facilities. It is a fact that these types of infrastructure are becoming more widespread but due to their variety and size, they cannot be directly utilized to make conclusions that help to improve the energy efficiency. ML approaches will be the key to the handling of energy efficiency problem in SBs. Learning about the occupants' consumption habits is capable of generating collaborative consumption predictions that help the occupant to consume better [258].

Some prominent projects in this space are described next. Reinisch et al. [259] developed an optimized application of AI system for SB environment. The system focuses on some capabilities like ubiquity, context awareness, conflict resolution, and self-learning features. The system operates on a knowledge base that stores all the information needed to fulfill the goals of energy efficiency and user comfort. Jahn et. al [260] proposed an energy efficiency features system built on top of a Hydra middleware framework [261]. The system provides both, stationary and mobile user interfaces for monitoring and controlling a smart environment. Pan et al. [262] proposed an IoT framework that uses smartphone platform and cloud-computing technologies to improve the energy efficiency in SBs. They built an experimental testbed for energy consumption data analysis. Fensel et al. [263] proposed the SESAME-S project (SEmantic SmArt Metering - Services for energy efficient houses). The project focuses on designing and evaluating the energy efficiency services to enable the end-consumers in making the right decisions and controlling their energy consumption. The system combines a variety of smart building components, such as smart meters, a variety of sensors, actuators, and simulators that can integrate virtual appliances such as the washing machine. Vastardis et al. [264] proposed a user-centric smart-home gateway system architecture to support home-automation, energy usage management, and smart-grid operations. The gateway is supported by ML classification algorithms component such as C4.5 and RIPPER that is able to extract behavioral patterns of the users and feed them back to the gateway.

Irrigation systems monitoring and smart watering system that keep track of rain and soil conditions and irrigate appropriately are a very cost-effective way to reduce outdoor water consumption. Investment in water management software and services, water-efficient plumbing, and irrigation management delivers economic and sustainability benefits. Water conservation and management is an example of such benefits [265].

*c) Comfort/Entertainment:* One of the main goals of SB research is to facilitate user daily life activities by increasing their satisfaction and comfort level. SBs supports automated appliance control and assistive services to offer a better quality of life. They utilize context awareness techniques to optimize the occupant's comfort based on predefined constraints of conditions in a building environment. Typical examples of comfort services include lighting, background music, automation of routine activities, advanced user interfaces based on voice or gestures, etc. [30]. Other services related to comfort services in SB environments are Indoor Climate Control and Intelligent Thermostat [265]. Indoor Climate Control: Measurement and control of temperature, lighting, CO2 fresh air. In the SB environment, HVAC systems play an essential role in forming indoor environmental quality. Typically, HVAC systems are produced not only to heat and cool the air but also to draw in and circulate outdoor air in large buildings [266]. Kabir et al. [267] present a context-aware application that provides the service according to a predefined preference of a user. They use the KNN classifier to infer the predefined service that will maximize the user's comfort and safety while requiring minimum explicit interaction of the user with the environment. Ahn et al. [268] proposed a deep learning model that estimates periodically the atmospheric changes and predict the indoor air quality of the near future.

*d) Safety/Security:* As the SB technology progresses, the role of ML and deep learning in security and connected devices will increase. Deep learning will continue to help gain insights using big data that were previously inaccessible, particularly in image and video. Advanced technologies such as behavioral analysis and ML to detect, categorize, and block new threats will be beneficial.

In a traditional home system, as soon as a fire is detected the Fire/smoke detectors are activated and start sending a fire alarm. However, SB can perform much better than the traditional system. It not only sends an alarm but also turns on the light only in the safest route and guides the occupants of the building out, as well as it will unlock the doors and windows for smoke ventilation, turn off all the devices and call the nearest fire service station. Other than this, it can take video of the areas surrounding the building, provide the status of window breakage alarms, and automatically lock all the doors and the windows when the last person of the house leaves [30].

The main services for security and safety in SBs are: Perimeter Access Control, Liquid Presence, Intelligent Fire Alarm, Intrusion Detection, and Motion Detection Systems [265]. Perimeter Access Control service provides control to restricted areas and detects non-authorized users that access the areas. Access card provides a variety of solutions that allow staff members, vendors or contractors to access specific areas at specific times you designate. The same access card can also be utilized to check employee attendance. In addition, there is widespread use of biometric technology including fingerprint, facial recognition, and iris scans [269]. Additionally, liquid presence detection technique has been utilized in data centers, warehouses, and sensitive building grounds to prevent breakdowns and corrosion in such areas [270].

Intelligent Fire Alarm and its corresponding safety systems are crucial parts of an intelligent building. It is a system with multi-function sensors (i.e., chemical gas sensors, integrated sensor systems, and computer vision systems) These sensors enable measuring smoke and carbon monoxide (CO) levels in the building. They also can give warnings, howling alarms, and tell with a human voice about the place and level of smoke and CO. In addition, they can give a message on a smartphone if the smoke or the CO alarm goes off [271]. Examples of intrusion detection systems including window and door opening detection and intrusion prevention [265]. An infrared motion sensor is utilized to detect the motion in a specific area in the building. This sensor can reliably send alerts to the alarm panel, with the system implementing algorithms for adaption to environmental disturbances and reducing any false alarms [265].

Image recognition solution can be used in security software to identify people, places, objects, and more. It can also be used to detect unusual patterns and activities. *Clarifai* [272] specializes in a field of ML known as "computer vision" that teaches computers to "see" images and video. Clarifai's technology can play a key role in security surveillance and at present, the company works only with home security. Each image is processed on a pixel by pixel basis through convoluted neural networks. Bangali and Shaligram [273] proposed a home security system that monitors the home when the user is away from the place. The system is composed of two methods: one uses a web camera to detect the intruder—whenever there is a motion detected in front of the camera, a security alert in terms of sound and an email is delivered to the occupant. And the other one is based on GSM technology that sends SMS. A home security system that sends alert messages to the house owner and police station in case of illegal invasion at home is proposed in [274]. The system consists of different sensor nodes as the input components while the output components respond to the signal received from the input components. The sensor nodes consist of a thief alarm, presence detecting circuit, and the break-in camera. Zhao and Ye [275] proposed a wireless home security system that utilizes low cost, low power consumption, and GSM/GPRS. The system has a user interface and it can respond to alarm incidents.

*e) Miscellaneous projects:* CASAS [178] is a project by Washington State University that provides a noninvasive assistive environment for dementia patients at SBs. The project focuses on three main areas for SBs: medical monitoring, green living, and general comfort. CASAS project comprises of three layers: physical layer, middleware layer, and software applications layer. *Aware Home Research Initiative (AHRI)* [276] is a project that has constructed by a group at the Georgia Institute of Technology for SB services in the fields of health and well-being, digital media and entertainment, and sustainability. AHRI utilizes a variety of sensors such as smart floor sensors, it also utilizes assistive robots for monitoring and helping the elderly.

*House_n* [277] is a multi-disciplinary project leads by a group of researchers at the MIT. The main objective of the project is to facilitate the design of the smart home and its associated technologies, products, and services. The home is supplied with hundreds of various sensors that are installed almost in every part of the home that and being utilized to develop user interface applications that enable the users to control and monitor their environment, save resources, remain mentally and physically active, and stay healthy.

The *EasyLiving* project [278] at Microsoft Research is concerned with the development of a prototype architecture and technologies to aggregate diverse devices into a coherent user experience for intelligent environments. The EasyLiving project was designed to provide context-aware computing services. The project utilizes a variety of sensors and cameras to track and recognize the human activities in the room by using the geometric model of a room and taking readings from sensors installed in the room.

The *Gator Tech Smart House* project [279] is a programmable space specifically designed for the elderly and disabled developed by The University of Florida's mobile and pervasive computing laboratory. The project's goal is to create smart building environments that can sense themselves and their residents. The project provides special cognitive services for the residents such as mobility, health, and other age-related impairments. A generic middleware is utilized to integrate system components in order to maintain a service definition for every sensor and actuator in the building. The components of the middleware including separate physical, sensor platform, service, knowledge, context management, and application layers [280].

Other well-known smart home projects include *DOMUS* [281] which is a research project, by the University of Sherbrooke in Canada, that supports mobile computing and cognitive assistance in smart buildings. The project aims to assist people suffering from Alzheimer's type dementia, schizophrenia, cranial trauma, or intellectual deficiencies.

*Adaptive House project* [136] at The University of Colorado has constructed a prototype system that is equipped with a variety of sensors that provide different environmental information including sound, motion, temperature, light levels. In addition, actuators that control the space and water heaters; lighting units, and ceiling fans.

In Asia, there are also some other smart building projects have been developed, such as "*Welfare Techno House*" project, which is equipped with different sensors such as ECG, body weight, and other temperature measured indicators [282]. *Ubiquitous Home* project [283] is another smart building project in Japan, which utilizes RFID, PIR, pressure sensors, as well as cameras and microphones for monitoring elderly adults.

*f) Summary:* Recently, several different context-aware and ML techniques have been utilized to support SB services. ML-based approaches are capable to perform better prediction and adaptation than others. The philosophy behind ML is to automate the learning process that enables algorithms to create analytical models with the support of available data. ML can be applied in different learning styles including supervised learning, unsupervised learning, semi-supervised learning, as well as reinforcement learning when the learning is the result of the interaction between a model and the environment. The general uses of ML for SB services are detection, recognition, prediction, and optimization. In the section, we also talked about how to acquire the context from multiple distributed and heterogeneous sources and the techniques for modeling and processing such context to be used in the application services of SBs. We also talked about the most used tools and platforms ML and others for real-time data analytics by ML community to efficiently process and learn from big data. Without such ML tools, one would have to implement all of the techniques from scratch requiring expertise in the techniques and in efficient engineering practices.

TABLE VIII: CATEGORIZED APPLICATIONS OF SB

| Application category | Cited | Characteristics | ML algorithm | Technology used |
|---|---|---|---|---|
| Elderly Population's Home Care | Chernbumroong et al. [56] | detection basic five activities namely feeding, grooming, dressing, mobility, and stairs. | SVM | wrist worn multi-sensors |
| | Taleb et al. [251] | Framework integrates both the sensing and the monitoring services for assisting elders at smart homes environment | NA | RFID readers with coverage of the whole house, video cameras, sound sensors, smart door lock, microphone and speakers |
| | CAALYX [252] | elderly monitoring subsystem, home monitoring subsystem and the caretaker's monitoring subsystem. | NA | vital sign sensors, GPS |
| | EasyLine+ [253] | support elderly people in carrying out a longer independent life at home. | neural network | illumination sensor, temperature sensor, door sensors, and RFID |
| Energy Efficiency | Reinisch et al. [259] | operates on a knowledge base that stores all information needed to fulfill the goals of energy efficiency and user comfort | AI methods | household appliances |
| | Jahn et. al [260] | stationary and mobile user interfaces for monitoring and controlling the smart environment | NA | wireless power metering plugs, household devices |
| | Fensel et al. [263] | designing and evaluating end consumer energy efficient services | NA | Smart meters, different types of sensors and actuators |
| | Vastardis et al. [264] | gateway system architecture to support home-automation, energy usage management, and smart-grid operations. | classification algorithms such as C4.5 and RIPPER | smart gateway |
| Safety and Security | Clarifai [272] | Computer vision platform for security surveillance in smart homes | CNN | surveillance cameras |
| | Bangali and Shaligram [273] | composed of two methods: web camera to detect the intruder, and GSM technology that sends SMS. | NA | web camera and GSM technology |
| | Zhao and Ye [275] | low cost, low power consumption | NA | GSM/GPRS |
| Comfort and entertainments | Kabir et al. [267] | provide service according to context-aware feature of the user | k nearest neighbors classifier | environment monitoring sensors |
| | Ahn et al. [268] | estimate the atmospheric changes and predict the indoor air quality | deep learning | carbon dioxide, fine dust, temperature, humidity, and light quantity sensors |
| Miscellaneous projects | CASAS [178] | medical monitoring, green living, and general comfort. | classification, regression and clustering algorithms. | Wearable sensors |
| | AHRI [276] | SB services in the fields of health and well-being, digital media and entertainment, and sustainability | NA | smart floor sensors, assistive robots |
| | House_n [277] | control people to control their environment, save resources, remain mentally and physically active | NA | Home environmental sensors |
| | EasyLiving project [278] | context-aware computing services through video tracking and recognition | NA | contains myriad devices that work together |

## VI. OPEN ISSUES AND FUTURE RESEARCH DIRECTIONS

Research on SBs has made great strides in recent years, but a number of challenges remain. We present some major challenges related to SBs in this part of the work. These challenges will channelize the research directions for future SBs.

### A. Security and Privacy

Wherever there is an interconnection of two systems or networks (wired or wireless), there are issues of security and privacy and the same is true in the case of SB. Security is an essential role in SB environments. Any SB application should ensure the confidentiality and integrity of data. Access control must be included in SB systems, for instance, the unauthorized users should not be able to disconnect the alarm system by connecting the pervasive system [284]. There is a massive amount of streaming that is collected from the various installed sensors and appliances, such data needs to be processed and stored. Hence, cloud computing services can be utilized for this purpose. However, with all of this data that is transmitted, the issue of losing the privacy increases. Therefore, different encryption techniques are needed to preserve personal privacy [285].

There are specific challenges related to the user's privacy including challenges related to the data privacy of personal information and the privacy of the individual's physical location and tracking. That needs for privacy enhancement technologies and relevant protection laws and tools for identity management of users and objects [286]. The recent trend of ML research has focused on handling security and privacy issues in SB environments. There are different security-related services have utilized ML techniques, such as determining safe device behavior by detecting and blocking activities and potentially harmful behavior [287].

ML techniques have the potential to reduce security gap because of their capability to learn, identify and detect the users' habits and behaviors. Consequently, it can detect the abnormal behaviors predicting risks and intrusions before they happen. For instance, ML models learn the routine of the users, such as the time they get home or go to sleep. These models can suggest rules based on those detected behaviors from all connected devices [288].

### B. SBs and context-aware computing

In the SB environment, there exists a massive amount of raw data being continuously collected about the various human activities and behaviors. It is important to develop techniques that convert this raw data into valuable knowledge [289]. Context awareness and ML techniques are expected to provide great support to process and store big data and create important knowledge from all this data [290].

The process of data interpretation and knowledge extraction has the following challenges including addressing noisy real-world data and the ability to develop further inference techniques that do not have the limitations of traditional algorithms. Usually, It is very complex to formalize and model the contextual information related to human behaviors in a standard way due to the complex physiological, psychological and behavioral aspects of human beings [291].

The humans communicate through rich languages as well as gestures and expressions. Modern ubiquitous computer systems lack an automatic mechanism of inferring information as the humans do. New research is necessary to raise human activities and behaviors recognition to understand the complex dependencies between the apps and humans [292], [293]. The context-aware prompting systems have essential applications in SBs such as emergency notifications, medication prompting, heart rate monitoring, generation of agenda reminders, and weather alerts. However, issuing prompts for all detected errors can possibly be false positives, and consequently, lead to annoyance and sometimes prove to be unsafe for specific activities. ML methods can be used for an accurate and precise prediction when a person faces difficulty while doing daily life activities [294].

*C. Personal Data Stream Management in SBs*

The data streaming management system is able to process and transfer raw data collected from a variety of sensors to information, it is also able to fuse this information to a feature and directly process features [295]. While the data processing for a single SB is simple, it is more complex when processing the data from multiple SBs, because there are different people that tend to share less common interests and have opposing interests concerning the processed data [296]. The simple sensors in an SB environment can detect different events related to temperature, motion, light, or weather. Moreover, other appliances like a television and a telephone can also send their status or other data as events. All this data from different sensors can be used by SB services to detect specific states and send a request to some actuators according to specific predefined rules, for instance, turn on the light if the television is used [297].

However, this approach is not generalizable in case of a group of people residing in the same building. Although it can work well for one certain person when personal preferences can be automatically learned for an individual person, therefore each of the residents has to define their own set of rules [298]. Because of the increasing number of sensors that produce data streams, the traditional analyzing and processing techniques of these data streams are mostly impractical now [299].

Despite the availability of new tools and systems for handling massive amounts of data continuously generating by a variety of sensors in SBs, however, the real promise of advanced data analytics to still lies beyond the realm of pure technology [296]. In [300] discusses research challenges for data streams of real-world applications. They analyze issues concerning privacy, timing, preprocessing, relational and event streams, model complexity and evaluation, availability of information, and problems related to legacy systems.

*D. Big data challenges in SBs*

Nowadays, a variety of sensing technology in the SBs can be utilized to collect a massive amount of heterogeneous data at a reasonable cost. Typically, hundreds of thousands of transactions can be generated by a single SB every day. The process of storing this data over the long-term is challenging [258]. We can imagine the challenges and opportunities that the companies or government will encounter in the future to manage incoming data from dozens of SBs. This new data could provide us with more contextual information that consequently leads to much better services to the occupants [301].

In the world of big data, despite the availability massive amount of data, however, it is not necessarily easy to obtain valuable information from this data utilizing the traditional approaches like trial and error to extract meaningful information from this data. Analyzing these massive amounts of data requires new technologies to store, organize, and process big data effectively, it needs high-performance processors that enable uncovering the insights in big data. It also requires flexible cloud computing services and virtualization techniques, as well as software such as Apache Hadoop and Spark [302]. It requires providing appropriate ML techniques which differ from the traditional approaches for effective and efficient solution of the above issues. For these reasons, researchers have recently started to think about the problems and opportunities resulting from the adoption of big data in SB environments [303] [304]. The information extracted from this big data has significant value and could greatly contribute in the future of SBs as assistive tools and for better services delivery. That is why it is necessary that the researchers start to analyze and think about the solutions for the current and future challenges of big data in SBs [305].

*E. Interoperability*

Interoperability means that two (or more) systems work together unchanged even though they were not necessarily designed to work together. When equipment, devices or appliances having different communication and networking technologies can communicate effectively, interoperability is satisfied. It is a challenge to ensure that an SB that has various components will be intelligible. Typically, each of these components might have been produced by different vendors, each of which may have created under different design constraints and considerations [306]. Therefore it becomes essential to satisfy interoperability so that a number of heterogeneous communication and networking technologies could coexist in various parts of SBs. For example, an energy management system may use Wi-Fi and ZigBee for communication purposes. A lot of work can be done in this context [307].

*F. Reliability*

We can expect that the reliability is one of the main concern of occupants and developers of SB systems. A variety of appliances and devices present in SB such as televisions, microwave, washing machines etc. are required exceedingly to be reliable. Achieving expected levels of reliability, especially when linked with communication technologies utilized with these devices that may be expected in SBs, is a great challenge. There are different reasons for these challenges differences in technological approaches, regulations, development culture, and the expectations of the market [306].

*G. Integration*

The key to a successful SB implementation is integration: linking building systems such as lighting, power meters, water meters, pumps, heating, and chiller plants together using sensors and control systems, and then connecting the building automation system to enterprise systems. Integration allows executives to gain smart-building benefits, both in new construction and by gradually transforming existing buildings into SBs. What these SBs have in common is integration. Generally, the integration in SB systems brings a range of benefits from energy savings to productivity gains to sustainability. The SB systems can be attached to enterprise business systems to add another level of intelligence that enhances decision-making and improves building performance [2].

However, integrating multiple systems is very challenging as each individual system has its own assumptions, strategies to control the physical world, and semantics. As an example of integrating two systems in SB, assume a system that is responsible for energy management, and another system for health care are running concurrently. In this case, the integrated system should not turn off medical appliances to save energy while they are being used as suggested by the health care system [292].

As a future perspective for SBs, You will wake up to the sound of the alarm, at the same time the available sensors will be aware that you are waking up. The other sensors such as light sensors will automatically turn on the light in the building, while the thermostat will warm the area that you are about to use in the building. Your coffee will start to brew, you will also get a notification on your phone about the weather. The other sensors in the kitchen and refrigerator will remind you with a list of items that you will need to pick up on your way from your workplace to home to make dinner. When you leave your house, you can press a button from your phone to self-drive your car out of the garage. After that, the security system will start monitoring and controlling the home. Such the doors will automatically lock. Appliances will switch to an energy-saving mode. When the home sensors sense utilizing geofencing technology that you are way back home, it will get ready again for your arrival, the thermostat will warm things up, the garage door will open as you pull up, and your favorite music will start to play when you walk in [141].

**Summary:** Although the recent researches have been done in the SBs field, there is a need for a lot more efforts; however, we believe that SBs are possible for the mass market in the near future. The main challenges and future research directions of this eld can be summarized as follows:

- User context in term of behavior and intention should be studied and respected whenever possible;
- Further research is needed into context-aware prompting systems, personal data streaming and big data analysis of occupants in SB environment;
- Some of the other challenges like the interoperability, reliability, and integration still require more attention.

VII. CONCLUSIONS

The promise of smart buildings (SBs) is a world of appliances that anticipate your needs and do exactly what you want them to at the touch of a button. Since SBs and their inhabitants create voluminous amounts of streaming data, SB researchers are looking towards techniques from ML and big data analytics for managing, processing, and gaining insights from this big data. This paper reviewed the most important aspects of SBs with particular focus on what is being done and what are the issues that require further research in ML and data analytics domains. In this regards, we have presented a comprehensive survey of the research works that relate to the use of ML and big data particularly for building smart infrastructure and services. Although the recent advancements in technologies that make the concept of SBs feasible, there are still a variety of challenges that limit large-scale real-world systems in SBs field. Addressing these challenges soon will be a powerful driving force for advancements in both industrial and academic fields of SB research.

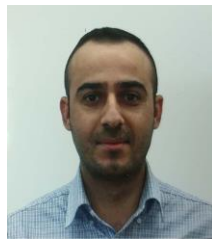

**Basheer Qolomany (S'17)** received the Ph.D. and second masters en-route to Ph.D. degrees in Computer Science from Western Michigan University (WMU), Kalamazoo, MI, USA, in 2018. He also received his B.Sc. and M.Sc. degrees in computer science from University of Mosul, Mosul, Iraq, in 2008 and 2011, respectively. He is currently an Assistant Professor at Department of Computer Science, Kennesaw State University, Marietta, GA, USA. Previously, he served as a Graduate Doctoral Assistant at Department of Computer Science, WMU, in 2016-2018; he also served as a lecturer at Department of Computer Science, University of Duhok, Kurdistan region of Iraq, Iraq, in 2011-2013. His research interests include machine learning, deep learning, Internet of Things, smart services, cloud computing, and big data analytics.

Dr. Qolomany has served as a Technical Program Committee (TPC) member and a reviewer of some international conferences include: IWCMC 2018, VTC 2018, MEDES 2016, and IC4 2016.

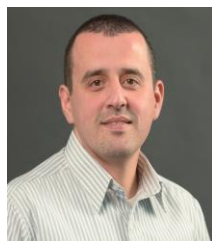

**Ala Al-Fuqaha (S'00-M'04-SM'09)** received Ph.D. degree in Computer Engineering and Networking from the University of Missouri-Kansas City, Kansas City, MO, USA, in 2004. His research interests include the use of machine learning in general and deep learning in particular in support of the data-driven and self-driven management of large-scale deployments of IoT and smart city infrastructure and services, Wireless Vehicular Networks (VANETs), cooperation and spectrum access etiquette in cognitive radio networks, and management and planning of software defined networks (SDN). He is a senior member of the IEEE and an ABET Program Evaluator (PEV). He serves on editorial boards and technical program committees of multiple international journals and conferences.

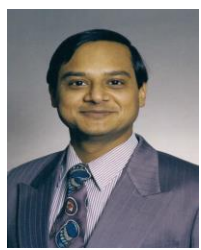

**Ajay Gupta (S'88 – M'89 – SM'05)** received his Ph.D. in Computer Science from Purdue University in 1989, his M.S. in Mathematics and Statistics from the University of Cincinnati in 1984, and his B.E. (Honors) in Electrical and Electronics Engineering from Birla Institute of Technology and Sciences, Pilani, India in 1982. He is currently a Professor of Computer Science at Western Michigan University, Kalamazoo, MI, USA. From 1998 to 2002, he was the Chairman of the Computer Science Department at Western Michigan University. He has also been two term Chair of the IEEE-CS Technical Committee on Parallel Processing from 2011 to 2015 and Vice-Chair of the Technical Activities Committee of the IEEE-CS in 2015-2016. His research interests include high performance computing, proteogenomics, data analytics, machine learning, sensor systems, cloud computing, mobile computing, web technologies, computer networks, evolutionary computation, scientific computing, and design and analysis of parallel and distributed algorithms. He has published numerous technical papers and book chapters in refereed conferences and journals in these areas.

Dr. Gupta is a senior member of the IEEE and member of the IEEE Computer Society, the IEEE Communications Society, the ASEE and the ACM. He actively helps organize various ACM and IEEE conferences. He is also involved in the global efforts to revise undergraduate and graduate computer science and computer engineering curriculum to keep pace with the technological advances.

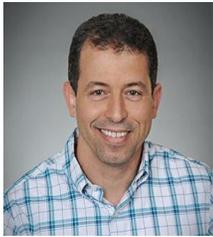**Driss Benhaddou (S'97 – M'02)** received the M.S. and two Ph.D. degrees in optoelectronics, engineering, and telecommunications from the University of Montpellier, France, and the University of Missouri-Kansas City, Kansas City, MO, USA, in 1991, 1995 and 2002, respectively. He is currently an Associate Professor and the Director of the Wireless and Optical Networking (WON) Research Laboratory, Department of Engineering Technology, University Houston, Houston, TX, USA. He served as the Principal Investigator (PI) or Co-PI on multiple research projects funded by NSF, NASA, Sprint, ATT, and the University of Houston. His research interests include Internet of Things applications to smart systems such as smart buildings, smart grid, smart cities, optical networking, sensor networks, switching system design, routing protocols, performance analysis, and optical instrument development for defect recognition of semiconductors.

Dr. Benhaddou has served as a Technical Program Committee Member and a reviewer of many international conferences and journals. He served as keynote speaker in many international conferences. He organized an NSF sponsored workshop on wireless application in smart cities in 2016 and co-chaired an IEEE conference on smart cities in 2017. He was the recipient of the Outstanding Researcher Award at the College of Technology, University of Houston in 2007.

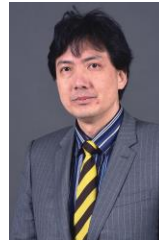**Alvis C. Fong (M'97 – SM'04)** received the BEng (Hons.) in information systems engineering and MSc in electrical engineering from Imperial College London, England, and PhD in electrical engineering from University of Auckland, New Zealand.

He began his professional career as a software RD Engineer with Motorola. Currently with Western Michigan University, MI, he has previously held faculty positions at Massey University, Nanyang Technological University, Auckland University of Technology, and University of Glasgow, as well as a visiting position at University of California Irvine. To date, he has published two books, 13 book sections, and more than 180 papers in leading international journals and conference proceedings, e.g. IEEE T-KDE, IEEE T-AC, IEEE T-II, IEEE T-EC, and contributed to two international patents owned by Motorola. His research interests are in applied AI and data mining for knowledge discovery.

Dr. Fong is a Fellow of IET, a Chartered Engineer registered in the UK, and a European Engineer. He has been an Associate Editor of IEEE T-CE since 2013.

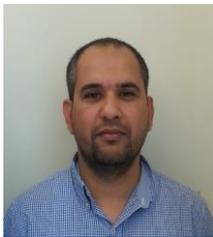**Safaa Alwajidi** received the BS and MS degrees in computer science from University of Baghdad, Iraq, in 2001 and 2004 respectively. He is currently working toward the Ph.D. degree with the Department of Computer Science, Western Michigan University (WMU), Kalamazoo, MI, USA. He is currently working as a part time instructor at Department of Computer Science, WMU. Mr. Alwajidi has been on the faculty of the Department of Computer Science at University of Baghdad since 2008. His research interests include big data visualization, algorithm design, machine learning and data mining.

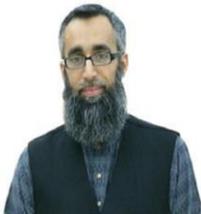**Junaid Qadir (M'14 – SM'14)** received Ph.D. from University of New South Wales, Australia in 2008 and his Bachelors in Electrical Engineering from UET, Lahore, Pakistan in 2000. He is an Associate Professor at the Information Technology University (ITU)–Punjab, Lahore since December 2015. He is the Director of the IHSAN (ICTD; Human Development; Systems; Big Data Analytics; Networks Lab) Research Lab at ITU (http://ihsanlab.itu.edu.pk/). Previously, he has served as an Assistant Professor at the School of Electrical Engineering and Computer Sciences (SEECS), National University of Sciences and Technology (NUST), from 2008 to 2015. His primary research interests are in the areas of computer systems and networking and using ICT for development (ICT4D).

Dr. Qadir has served on the program committee of a number of international conferences and reviews regularly for various high-quality journals. He is an Associate Editor for IEEE Access, Springer Nature Central's Big Data Analytics journal, Springer Human-Centric Computing and Information Sciences, and the IEEE Communications Magazine. He is an award-winning teacher who has been awarded the highest national teaching award in Pakistanthe higher education commissions (HEC) best university teacher awardfor the year 2012-2013. He has considerable teaching experience and a wide portfolio of taught courses in the disciplines of systems networking; signal processing; and wireless communications and networking. He is a member of ACM, and a senior member of IEEE.